\documentclass[a4paper,fleqn,usenatbib]{mnras}

\usepackage[T1]{fontenc}
\usepackage{ae,aecompl}
\usepackage{textcomp}

\usepackage{graphicx}	
\usepackage{amsmath}	
\usepackage{amssymb}	
\usepackage{multicol}        
\usepackage{bm}		
\usepackage{pdflscape}	
\usepackage{color}          
\title[Structure of the SMC from RR Lyrae stars]{The VMC survey - XXVI. Structure of the Small Magellanic Cloud from RR Lyrae stars\thanks{Based on observations made with VISTA at ESO under programme ID 179.B-2003.}}

\author[T. Muraveva et al.]{T. Muraveva$^{1}$\thanks{tatiana.muraveva@oabo.inaf.it}, S. Subramanian$^{2}$, G. Clementini$^{1}$, M.-R.L. Cioni$^{3,4}$, M. Palmer$^{5}$,
\newauthor{J.Th. van Loon$^{6}$, M.~I. Moretti$^{7}$, R. de Grijs$^{2,8,9}$, R. Molinaro$^{7}$, V. Ripepi$^{7}$,}
\newauthor{M. Marconi$^{7}$, J.~Emerson$^{10}$, V.~D.~Ivanov$^{11,12}$}
\\
$^{1}$ INAF-Osservatorio Astronomico di Bologna, Via Gobetti, 93/3, Bologna 40129, Italy\\
$^{2}$ Kavli Institute for Astronomy and Astrophysics, Peking University, Yi He Yuan Lu 5, Hai Dian District, Beijing 100871, China\\
$^{3}$ Leibnitz-Institut f\"{u}r Astrophysik Potsdam, An der Sternwarte 16, 14482 Potsdam, Germany\\
$^{4}$ University of Hertfordshire, Physics Astronomy and Mathematics, College Lane, Hatfield AL10 9AB, UK\\
$^{5}$ Dept. d'Astronomia i Meteorologia, Institut de Ci\`encies del Cosmos, Universitat de Barcelona (IEEC-UB), Mart\'i Franqu\`es 1,\\ E-08028 Barcelona, Spain\\ 
$^{6}$ Lennard-Jones Laboratories, Keele University, ST5 5BG, UK\\
$^{7}$ INAF-Osservatorio Astronomico di Capodimonte, via Moiariello 16, Naples 80131, Italy \\
$^{8}$ Department of Astronomy, Peking University, Yi He Yuan Lu 5, Hai Dian District, Beijing 100871, China\\
$^{9}$ International Space Science Institute--Beijing, 1 Nanertiao, Zhongguancun, Hai Dian District, Beijing 100190, China\\
$^{10}$ Astronomy Unit, School of Physics and Astronomy, Queen Mary University of London, Mile End Road, London E1 4NS, UK\\
$^{11}$ European Southern Observatory, Ave. Alonso de C{\'o}rdova 3107, Vitacura, Santiago, Chile\\
$^{12}$ European Southern Observatory, Karl-Schwarzschild-Strasse 2, D-85748 Garching bei M\"{u}nchen, Germany\\
}

\date{Accepted . Received ; in original form }

\pubyear{2016}

\begin{document}
\label{firstpage}
\pagerange{\pageref{firstpage}--\pageref{lastpage}}
\maketitle

\begin{abstract}
We present  results from the analysis of 2997 fundamental mode RR Lyrae variables located in the Small Magellanic Cloud (SMC). For these objects near-infrared time-series photometry from the {\it VISTA survey of the Magellanic Clouds system} (VMC) and visual light curves from the OGLE~IV survey are available. In this study the multi-epoch  $K_{\rm s}$-band VMC photometry was used for the first time to derive  intensity-averaged magnitudes of the SMC RR Lyrae stars.  We determined individual distances to the RR Lyrae stars from the near-infrared  period-absolute magnitude-metallicity ($PM_{K_{\rm s}}Z$) relation, which has a number of advantages in comparison with the visual absolute magnitude-metallicity ($M_{V}-{\rm [Fe/H]}$) relation, such as a smaller dependence of the luminosity on interstellar extinction, evolutionary effects and metallicity. The distances we have obtained were used to study the three-dimensional structure of the SMC. The distribution of the SMC RR Lyrae stars is found to be ellipsoidal.  The actual line-of-sight depth of the SMC is in the range from 1 to 10~kpc, with an average depth of 4.3 $\pm$ 1.0 kpc. We found that RR Lyrae stars in the eastern part of the SMC are affected by interactions of the Magellanic Clouds.  However, we do not see a clear bimodality  in the distribution of RR Lyrae stars as observed for red clump (RC) stars.  

\end{abstract}

\begin{keywords}
Surveys -- stars: variables: RR Lyrae -- galaxies: Magellanic Clouds -- galaxies: structure
\end{keywords}


\section{Introduction}
The Small Magellanic Cloud (SMC) is a nearby  dwarf irregular galaxy. It is part of the Magellanic System (MS) which also comprises the Large Magellanic Cloud (LMC), the Magellanic Bridge (MB) and the Magellanic Stream.
The SMC gravitationally interacts with the LMC and the Milky Way (MW). As a result it has a complex internal structure characterised by a disturbed shape and a large extent along the line-of-sight \citep{Putman1998}.

%

RR Lyrae stars are old (age $>10$ Gyr),  low-mass ($\sim0.6 - 0.8$ $M_\odot$), radially-pulsating variables located on the horizontal branch of the colour-magnitude diagram (CMD).
They pulsate in the fundamental mode (RRab), first-overtone mode (RRc) or both modes  simultaneously (RRd).  RR Lyrae stars are abundant in globular clusters and in the halos of galaxies. They  could serve to study the structure, interaction history and the distance to the parent systems because they follow an absolute magnitude-metallicity (${M_{V}-{\rm [Fe/H]}}$) relation 
in the visual band and infrared period-luminosity ($PL$) and $PL$-metallicity ($PLZ$) relations.

The three-dimensional structure of the SMC as traced by RR Lyrae stars has been the subject of a number of studies (e.g. \citealt{Grah1975}; \citealt{Sos2010};  \citealt{Sub2012}; \citealt{Has2012}; \citealt{Kap2012}; \citealt{Deb2015}; \citealt{Jac2016}, \citealt{Deb2017}). \citet{Grah1975} analysed 76 SMC RR Lyrae stars finding that these objects are distributed smoothly and do not show a strong concentration in the bar or in the centre of the SMC. 
\citet{Sos2010} published the catalogue of RR Lyrae stars observed in the SMC  by the third phase of the Optical Gravitational Lensing Experiment (OGLE~III) and concluded that the distribution of RR Lyrae stars in the galaxy is roughly round on the sky with two maxima near the centre. 
\citet{Sub2012} analysed the relative positions of different regions of the SMC inferred from the $V$ and $I$ photometry of the RR Lyrae variables observed by  OGLE~III. According to this study the SMC RR Lyrae stars have an ellipsoidal distribution and the north-eastern part of the SMC is located  closer to us. Similarly, \citet{Has2012} used the SMC RR Lyrae stars  from the OGLE~III survey and found that the RR Lyrae stars show a spheroidal or ellipsoidal distribution with an off-centered and nearly bimodal  peak.  \citet{Deb2015} found that the north-eastern arm of the SMC is  located 
closer than the plane of the SMC main body. 
These authors  studied the depth along the line-of-sight and concluded that it is larger for the central part of the SMC. \citet{Kap2012} studied a sample of RRab stars located in 14 ${\rm deg^2}$ of the SMC and found that the north-eastern part of the SMC has a greater depth. Furthermore, they studied  the metal abundance and the spatial distribution of the RR Lyrae variables and suggested that metal-richer and metal-poorer objects in the sample belong to different dynamical structures. \citet{Jac2016} studied the RR Lyrae stars using OGLE~IV data, which covers almost entirely the SMC. They confirm the ellipsoidal shape of the SMC and do not find any sub-structures. However, they detect some asymmetry in the equal density contours in the eastern part of the SMC.  Namely, the center of the ellipsoid shifts to east and towards the observer.  More recently, \citet{Deb2017} used the sample of the OGLE~IV RRab stats to determine the distance, reddening and structural parameters of the SMC.

In all studies mentioned above visual photometry was used to analyse the SMC RR Lyrae stars. In the present  study, for the first time,  $K_{\rm s}$-band multi-epoch photometry from  the {\it VISTA survey of the Magellanic Clouds system} (VMC) for 2997 RRab stars distributed across a large area  ($\sim$42~${\rm deg^2}$) of the SMC, was used to probe the galaxy's structure. 
The near-infrared ($K$ or $K_{\rm s}$ bands) period-absolute magnitude-metallicity  ($PM_KZ$) relation has a number of advantages in comparison with the optical ${M_{V}-{\rm [Fe/H]}}$ relation,  specifically,  a smaller dependence of the luminosity on interstellar extinction ($A_K$=0.114$A_V$), evolutionary effects and metallicity. It was discovered by \citet{Long1986} and later studied by different authors from both a theoretical and an observational point of view (e.g. \citealt{Bono2003};  \citealt{Cat2004};  \citealt{diCr2004}; \citealt{DelP2006}; \citealt{Sol2006,Sol2008}; \citealt{Bor2009}; \citealt{Marconi2015}; \citealt{Mur2015}). 
Here we apply the relation from \citet{Mur2015} to derive individual distances to RR Lyrae stars in the SMC sample and study the structure of this galaxy.

 In Section~\ref{sec:dat} we provide information about our sample of RR Lyrae variables in the SMC and the data available for these objects from the 
 VMC and OGLE~IV surveys.
 In Section~\ref{sec:pl} we analyse  the $PK_{\rm s}$ relations for the whole sample of SMC RR Lyrae stars as well as for individual VMC tiles. 
  We present the three-dimensional structure of the SMC as traced by RR Lyrae stars in Section~\ref{sec:str}.  Finally, Section~\ref{sec:sum} provides a summary of our results and main conclusions. 

\section{Data}\label{sec:dat}
\subsection{The VMC survey\label{sec:vmc}}
VMC \citep{Cioni2011} is an ongoing imaging survey of the MS in the $Y$, $J$, $K_{\rm s}$ passbands, centred at $\lambda$ = 1.02, 1.25 and 2.15 $\mu$m, respectively. 
 It started in 2009 and observations of the MS were 89\% complete as of  September 2017, while the observations of the SMC were 100\% complete.  The survey covers the LMC area ($\sim$105~$\rm deg^2$) with 68 tiles, the SMC ($\sim$42~$\rm deg^2$) with 27 tiles, the MB area ($\sim$21~$\rm deg^2$) with 13 tiles and part of the Stream ($\sim$3~$\rm deg^2$) with 2 tiles.  The VMC $K_{\rm s}$-band  observations are taken over 13 separate epochs: 11 times with an exposure time of 750~s (deep epochs) and twice with an exposure time of 375~s (shallow epochs). Additional epochs may be obtained if observations have to be repeated because the requested sky conditions are not met \citep{Cioni2011}. 
Every single deep epoch reaches a limiting magnitude of $K_{\rm s} \sim 19.2$~mag with a signal-to-noise ratio $S/N = 5$, in the Vega system. VMC reaches a sensitivity limit on the stacked images of $K_{\rm s}=21.5$~mag with  $S/N = 5$.
 The VMC images are processed by the Cambridge Astronomical Survey Unit (CASU; \citealt{Lew2010}). The data are then sent to the Wide Field Astronomy Unit (WFAU) in Edinburgh where the single epochs are stacked, catalogued and ingested into the  VISTA Science Archive (VSA;  \citealt{Cross2012}). 
 
 The strategy, main science goals and the first data from  the VMC survey were described in \citet{Cioni2011}. Analysis of variable stars based on VMC data was presented in \citet{Rip2012a,Rip2012b}, \citet{Ripepi2016}, \citet{Moretti2016} and \citet{Marconi2017} for classical Cepheids, \citet{Rip2014} for Anomalous Cepheids, \citet{Rip2015} for Type II Cepheids, in \citet{Mor2014} for classical Cepheids, RR Lyrae stars and eclipsing binaries (EBs), in \citet{Mur2015} for RR Lyrae stars and in \citet{Mur2014} for EBs.  Using the VMC observations of the red clump (RC) stars  \citet{Sub2017} and \citet{Tatton2013} studied the structure of the SMC and the 30~Doradus region in the LMC, respectively.  \citet{Rub2015} analysed the star formation history of the SMC using the VMC photometry.
\begin{table*}
	\centering
\caption{ VMC tiles in the SMC analysed in the present study: (1) Field and tile number; (2), (3) Coordinates of the tile centre J2000; (4) Number of epochs in the $K_{\rm s}$ band,  available at the moment of analysis including observations obtained in nights with sky conditions that did not meet the VMC requirements (see text for details); (5) Number of RRab stars; (6)  $PK_{\rm s}$ relation of RR Lyrae stars in the tile; (7) r.m.s. of the relation; (8) Distance moduli of the tiles.}
\label{tab:tiles}
\begin{tabular}{@{}ccccccccc@{}}
\hline 
Tile  & $\alpha$ & $\delta$ & N & ${\rm N_{RR\,Lyr}}$ & $PK_{\rm s}$ & r.m.s. &$(m-M)_0$ \\
 & ($^{\rm h}:^{\rm m}:^{\rm s}$) & ($^\circ :^\prime:^{\prime\prime}$) & epochs &  &   relation  & (mag)  & (mag) \\
\hline 
\hline
SMC 2\_2 & 00:21:43.920 & $-$75:12:04.320 &   12 &  67 & $(-3.03\pm0.46)\log P + (17.71\pm0.10)$ & 0.15 & $18.87\pm0.18$  \\
SMC 2\_3 & 00:44:35.904 & $-$75:18:13.320 &   15 &  103 & $(-3.02\pm0.46)\log P + (17.71\pm0.10)$ & 0.16 & $18.89\pm0.18$  \\
SMC 2\_4 & 01:07:33.864 & $-$75:15:59.760 &   15 &   93  & $(-2.89\pm0.39)\log P + (17.72\pm0.09)$ &  0.14 & $18.86\pm0.15$ \\
SMC 2\_5 & 01:30:12.624 & $-$75:05:27.600 &   16 &   76  & $(-2.01\pm0.60)\log P + (17.89\pm0.14)$ &  0.19 & $18.84\pm0.20$ \\
SMC 3\_1 & 00:02:39.912 & $-$73:53:31.920 &   16 & 26 & $(-4.14\pm0.54)\log P + (17.43\pm0.12)$ & 0.13 & $18.82\pm0.17$  \\
SMC 3\_2 & 00:23:35.544 & $-$74:06:57.240 &   14 & 102 & $(-2.96\pm0.36)\log P + (17.75\pm0.08)$ & 0.14 & $18.88\pm0.19$  \\
SMC 3\_3 & 00:44:55.896 & $-$74:12:42.120 &   18  & 207 & $(-3.41\pm0.28)\log P + (17.64\pm0.07)$ & 0.16 & $18.91\pm0.18$  \\
SMC 3\_4 & 01:06:21.120 & $-$74:10:38.640 &   16  &  190 & $(-3.45\pm0.27)\log P + (17.60\pm0.06)$ & 0.16 & $18.86\pm0.17$   \\
SMC 3\_5 & 01:27:30.816 & $-$74:00:49.320 &   16  &  96 & $(-2.33\pm0.41)\log P + (17.83\pm0.09)$ & 0.17 & $18.85\pm0.19$  \\
SMC 3\_6 & 01:48:06.120 & $-$73:43:28.200 &   16  & 38 & $(-2.28\pm0.80)\log P + (17.77\pm0.18)$ & 0.19 & $18.78\pm0.20$   \\
SMC 4\_1 & 00:05:33.864 & $-$72:49:12.000 &   13 &   41 & $(-2.05\pm0.64)\log P + (17.95\pm0.14)$ & 0.14  & $18.85\pm0.26$   \\
SMC 4\_2 & 00:25:14.088 & $-$73:01:47.640 &   15 &  142  & $(-3.22\pm0.29)\log P + (17.72\pm0.07)$ & 0.13 & $18.94\pm0.19$  \\
SMC 4\_3 & 00:45:14.688 & $-$73:07:11.280 &   16 & 279  & $(-3.57\pm0.31)\log P + (17.61\pm0.07)$ & 0.24 & $18.90\pm0.27$  \\
SMC 4\_4 & 01:05:19.272 & $-$73:05:15.360 &   15 & 250  & $(-4.12\pm0.24)\log P + (17.48\pm0.06)$ & 0.17 & $18.88\pm0.21$  \\
SMC 4\_5 & 01:25:11.088 & $-$72:56:02.760 &   18 &  124 & $(-2.20\pm0.50)\log P + (17.86\pm0.11)$ & 0.19 & $18.84\pm0.21$  \\
SMC 4\_6 & 01:44:34.512 & $-$72:39:44.640 &   16 &  42 & $(-3.13\pm0.82)\log P + (17.64\pm0.19)$ & 0.18 & $18.85\pm0.18$  \\
SMC 5\_2 & 00:26:41.688 & $-$71:56:35.880 &   17 &  89 & $(-3.01\pm0.45)\log P + (17.74\pm0.10)$ & 0.14 & $18.88\pm0.19$  \\
SMC 5\_3 & 00:45:32.232 & $-$72:01:40.080 &   19 &  174 & $(-3.32\pm0.29)\log P + (17.67\pm0.06)$ & 0.15 & $18.91\pm0.16$  \\
SMC 5\_4 & 01:04:26.112 & $-$71:59:51.000 &   18 & 188 & $(-3.02\pm0.34)\log P + (17.72\pm0.08)$ & 0.20 & $18.89\pm0.23$  \\
SMC 5\_5 & 01:23:09.336 & $-$71:51:09.720 &   14 & 111  & $(-3.57\pm0.45)\log P + (17.56\pm0.10)$ & 0.17 & $18.82\pm0.21$  \\
SMC 5\_6 & 01:41:28.800 & $-$71:35:47.040 &   18 & 53 & $(-3.30\pm0.58)\log P + (17.60\pm0.13)$ & 0.14 & $18.83\pm0.14$  \\
SMC 6\_2 & 00:28:00.192 & $-$70:51:21.960 &   17 &  83 & $(-1.57\pm0.45)\log P + (18.05\pm0.10)$ & 0.14 & $18.90\pm0.16$  \\
SMC 6\_3 & 00:45:48.792 & $-$70:56:09.240 &   14 & 120 & $(-3.01\pm0.30)\log P + (17.72\pm0.07)$ & 0.14 & $18.87\pm0.17$  \\
SMC 6\_4 & 01:03:40.152 & $-$70:54:25.200 &   13 &  100 & $(-3.00\pm0.31)\log P + (17.72\pm0.07)$ & 0.14 & $18.87\pm0.17$  \\
SMC 6\_5 & 01:21:22.560 & $-$70:46:11.640 &   14 & 71  & $(-3.06\pm0.46)\log P + (17.70\pm0.10)$ & 0.14 & $18.86\pm0.19$  \\
SMC 7\_3 & 00:46:04.728 & $-$69:50:38.040 &   16 &  56 & $(-2.70\pm0.56)\log P + (17.81\pm0.13)$ & 0.14 & $18.91\pm0.16$   \\ 
SMC 7\_4 & 01:03:00.480 & $-$69:48:58.320 &   15 &  76 & $(-3.37\pm0.47)\log P + (17.63\pm0.11)$ & 0.14 & $18.88\pm0.15$   \\ 

\hline
\end{tabular}
\end{table*}

\begin{figure}
   \includegraphics[trim=30 180 30 100,width=1.05\linewidth]{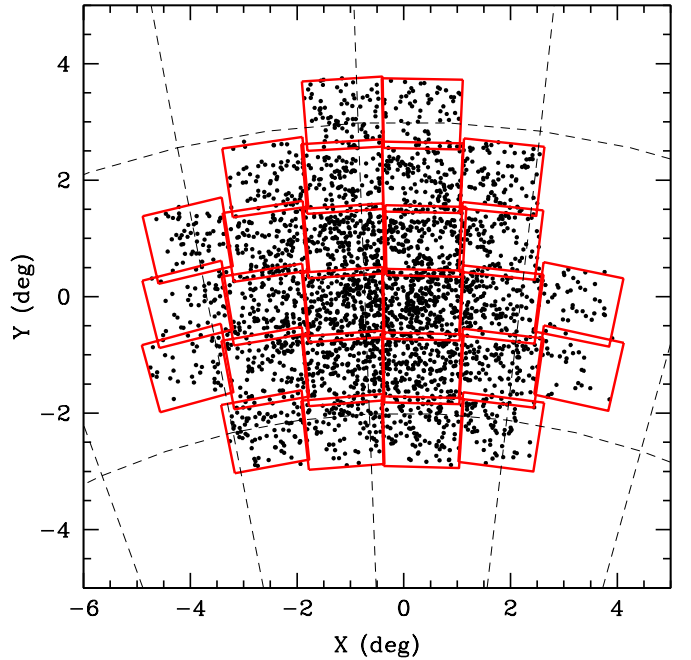}
  \caption{VMC coverage of the SMC.  Red boxes represent  27 VMC tiles.  Black dots mark the 2997 RR Lyrae stars analysed in this paper. Coordinates are defined as in \citet{vandM2001} where $\alpha_0$ = 12.5 deg, $\delta_0$ = $-$73 deg.}
  \label{fig:map}
\end{figure}

To study the $K_{\rm s}$-band light curves of the SMC RR Lyrae stars we used VMC observations in all 27 SMC tiles. We analysed the $K_{\rm s}$-band light curves with the GRaphical Analyser of TImes Series (GRATIS), custom software developed at the Observatory of Bologna by P. Montegriffo (see e.g. \citealt{Clem2000}). This software requires at least 11  epochs to model the light curves and properly compute intensity-averaged magnitudes. However, even though all sources located in the analysed tiles are expected to have 13 good quality epochs, stars  affected by some problems, such as blending, may not necessarily be detected in all epochs and, thus, have the 11 data points necessary for the analysis with GRATIS.  Hence, initially we used all available VMC epochs including observations obtained in nights with sky conditions (seeing and ellipticity)  that did not meet the VMC requirements \citep{Cioni2011} and discarded data points that deviated significantly from the model fit during the analysis with GRATIS. For some stars the model line fits well all data points, hence, we used all available data in the analysis. The VMC coverage of the SMC is shown in Figure~\ref{fig:map}, where red boxes represent the 27 SMC tiles ($\sim42$~${\rm deg^2}$) used in this study. The $X$ and $Y$ axes in Fig.~\ref{fig:map} are the coordinates of a zenithal equidistant projection as defined by \citet{vandM2001}.

\begin{figure*}
  \includegraphics[trim=50 170 50 100,width=16cm]{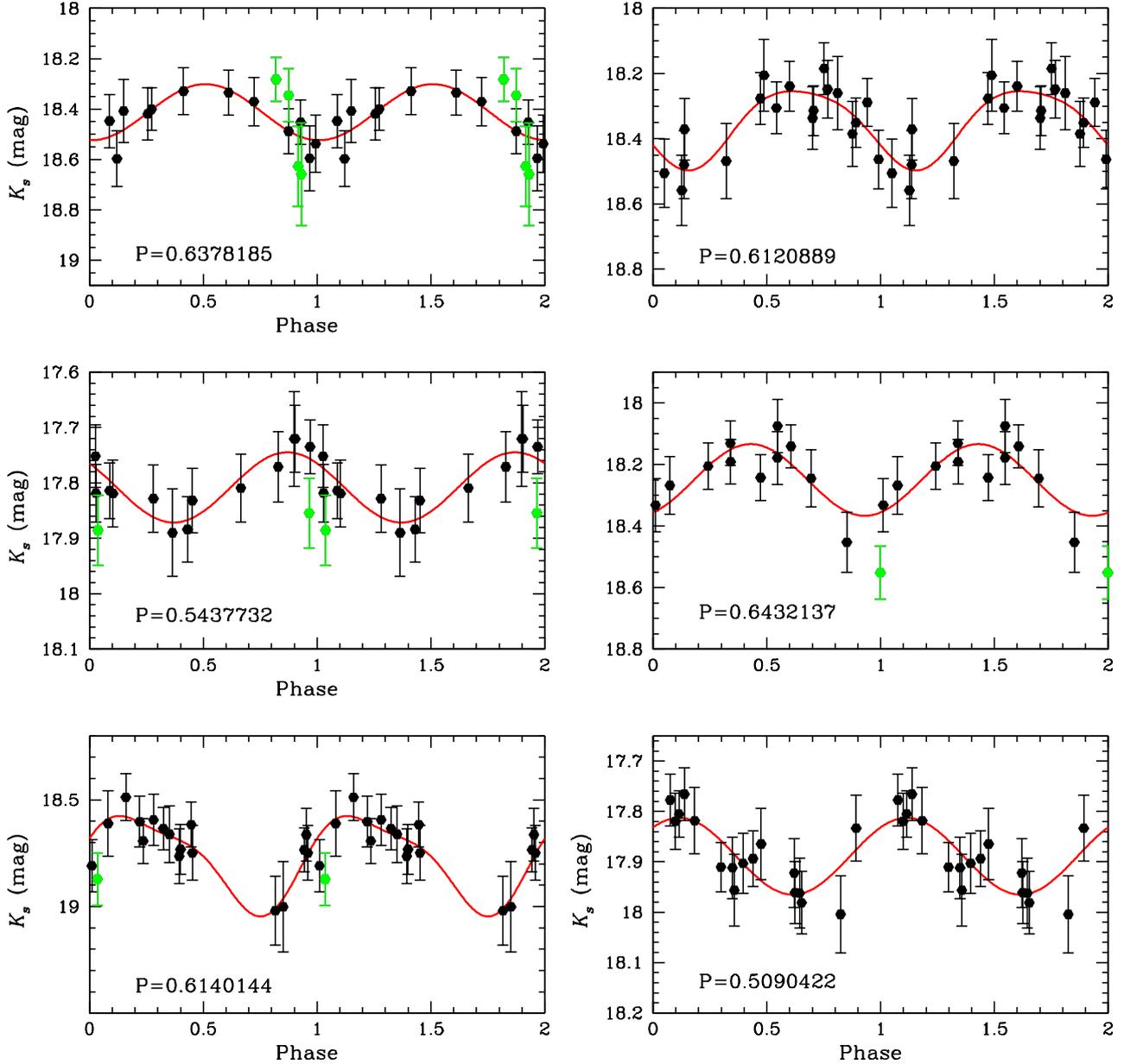}
  \caption{Phased $K_{\rm s}$-band light curves of the SMC RRab stars observed by the VMC survey with 11 or more epochs. Periods are from the OGLE~IV catalogue and measured in days.  Black and green dots represent data points used and discarded from the analysis, respectively. The red line shows the best fit model obtained with GRATIS. \label{fig:lc}}
  \label{}
\end{figure*}
The time sampling of the VMC survey and the significantly reduced amplitudes of the light variations in the $K_{\rm s}$ passband allow us to determine mean $K_{\rm s}$ magnitudes for RR Lyrae and other pulsating stars with great precision (\citealt{Rip2012a,Rip2014}). On the other hand,  the small amplitude of pulsation in the $K_{\rm s}$ band complicates the search for new variables  based solely on near-infrared data.  A method to identify variable stars in the MS based only on the VMC photometry was developed by \citet{Moretti2016}, but it is more suited to search for classical Cepheids. Hence, our study is based only on known SMC variable stars identified by optical microlensing surveys such as OGLE~IV \citep{Sos2016}. 

\subsection{OGLE~IV}
The initial goal of the OGLE survey was the search for microlensing events, but as a byproduct the survey also discovered a large number of variable stars in the MS and in the MW bulge. The OGLE~IV catalogue comprises observations of about 650 $\rm deg^2$ in the MS obtained between  March 2010 and July 2015 with the 1.3 m Warsaw telescope at the Las Campanas Observatory, Chile \citep{Udalski2015}.  Observations were performed in the Cousins $I$ passband with the number of data points ranging from 100 to 750 and  in the Johnson $V$ band with the number of observations ranging from several to 260.  The OGLE~IV catalogue is publicly available from the OGLE website\footnote{http://ogle.astrouw.edu.pl} and contains, among others, 45451 RR Lyrae variables, of which 6369 stars are located towards the SMC \citep{Sos2016}. The catalogue provides right ascension (RA), declination (Dec), mode of pulsation, mean $V$ and $I$ magnitudes, period of pulsation, $I$-band amplitude, parameters of the Fourier decomposition and  time-series $V,I$ photometry.

Among the SMC RR Lyrae variables we selected only RRab stars for our analysis, since RRc and RRd stars have smaller amplitudes and noisier light curves, hence, it is particularly complicated to fit their near-infrared  light curves and obtain reliable mean $K_{\rm s}$ magnitudes. 
OGLE~IV provides information on 4961 RRab stars in the SMC, of which 3484 variables have counterparts in the VMC catalogue within 1\arcsec. Among 1477 variables that do not have VMC counterpart within 1\arcsec, 1336 are located outside the VMC footprint, while 141 stars are in the VMC footprint but we did not find them likely because of coordinates' uncertainty.  Adopting a cross-matching radius of 2 arcsec instead of 1 arcsec we were able to recover  VMC counterparts for  100 among 141 stars, and 125 stars are recovered  within a radius of 5 arcsec. However,  it is risky to add these stars to the sample, since wrong crossmatches could happen.  Stars  that do not have a VMC counterpart even within 5 arcsec are aligned along the edges of the VMC footprint, located in the gap between VMC tiles 5\_3 and 5\_4 or are clustered in some kind of holes of the VMC coverage likely caused by saturated  MW stars in front of the SMC. However, the percentage of objects that we lose owing to these issues is rather small. 

  We discarded from the sample of 3484 variables which  have counterparts in the VMC catalogue within 1\arcsec, stars that have fewer than 11 $K_{\rm s}$ band epochs  (see Section~\ref{sec:vmc}) and sources observed by VISTA detector 16. Detector 16 is affected by a time-varying quantum efficiency which makes accurate flat fielding impossible \citep{Jarvis2013}. This effect is more significant in the shorter wavelength filters but also present in the $K_{\rm s}$  band. After these cleaning procedures we are left with a sample of  3121 RR Lyrae variables.

 We analysed the $K_{\rm s}$-band light curves of the  3121 RR Lyrae stars with GRATIS,  and derived intensity-averaged  magnitudes and amplitudes using the periods provided by OGLE~IV. Examples of the $K_{\rm s}$-band light curves are shown in Fig.~\ref{fig:lc}, where  black and green dots represent the data points used and discarded from the analysis, respectively. Red lines represent the best fits obtained with GRATIS.
 
 \begin{figure}
  \includegraphics[trim=20 170 20 90 clip,width=\linewidth]{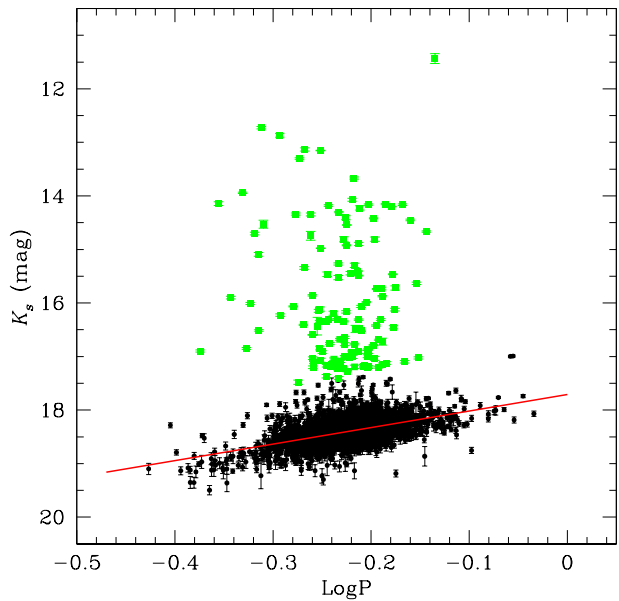}
  \caption{$PK_{\rm s}$ relation of  3121 SMC RRab stars from the OGLE~IV catalogue which have a counterpart and 11 or more $K_{\rm s}$-band epochs in VMC and were not observed by VISTA detector 16. The red line represents the best non-weighted least squares fit. Black circles and green squares are stars located within and beyond $5\sigma$ from the best fit line, respectively.}
  \label{fig:PL_all_MW}
\end{figure}

 The distribution of the  3121 RR Lyrae stars in the $K_{\rm s}$ magnitude versus  logarithmic period   plane is shown in Fig.~\ref{fig:PL_all_MW}. The large majority of the stars  follow the $PK_{\rm s}$ relation. However, there is a group of objects that do not follow the $PK_{\rm s}$ relation and have significantly brighter magnitudes. They are likely MW members or sources blended with close companions. A division between SMC and MW RR Lyrae stars is not provided in the OGLE~IV catalogue since  \citet{Sos2016} pointed out  that it was impossible to separate the MW and SMC old stellar populations owing to the interaction between the two galaxies. However, in the current analysis we want to focus on the SMC inner structure. Hence, we perform an approximate separation between the two populations and clean the sample from blended sources.
 In order to do this we perform  a linear non-weighted least squares fit  to the whole sample of  3121 RR Lyrae variables (red line in Fig.~\ref{fig:PL_all_MW}) by progressively discarding objects which deviate more than $5\sigma$ from the best fit line (122 sources in total, shown by  green squares in Fig.~\ref{fig:PL_all_MW}).  
 
We chose a rather large scatter ($5\sigma$ from the $PK_{\rm s}$ relation) to distinguish objects that belong to the SMC because the actual dispersion of the $PK_{\rm s}$ relation in Fig.~\ref{fig:PL_all_MW} includes not only the intrinsic dispersion of the relation, but also the scatter caused by the different distances spanned by the RR Lyrae stars in our  sample (depth effect). Selecting stars located within a smaller interval, for example  $3\sigma$, is risky, since it causes the removal of RR Lyrae variables that actually belong to the SMC but are significantly scattered from the $PK_{\rm s}$ owing to the large extension of the SMC along the line-of-sight.

 Of the  122 stars that scatter more than $5\sigma$ from the $PK_{\rm s}$ relation  72 are too bright also in the OGLE~IV catalogue and significantly scattered from the $PL$ relation in $I$ band, hence, they are either MW members or stars blended in both, the $I$ and $K_{\rm s}$ passbands. We discarded them from the following analysis. Additional  50 objects have $V$ and $I$ magnitudes consistent with the distance to the SMC, but too bright magnitudes in the $K_{\rm s}$ passband. We checked  their OGLE and VMC images and found that all of them have close companions  and are blended in the $K_{\rm s}$ band, hence should be discarded.  We are thus left with a sample of  2999 RR Lyrae stars. Classification of one of them  (OGLE-SMC-RRLYR-1505) was marked as uncertain in the OGLE~IV catalogue.  Another object (OGLE-SMC-RRLYR-3630) represents two RRab stars with different periods located on the line of sight. We analysed the $K_{\rm s}$-band light curve of this star using both periods but we were not able to distinguish which of the two periods is the correct one  owing to the small number of data points in the $K_{\rm s}$ band. After discarding stars  OGLE-SMC-RRLYR-1505 and OGLE-SMC-RRLYR-3630 our final sample comprises 2997 RR Lyrae variables, which are shown as black dots in Fig.~\ref{fig:map}.  Their main properties are presented in Table~\ref{tab:gen}.

\subsection{Extinction}\label{sec:ext}


The reddening in the SMC has been studied by several authors  (e.g. \citealt{Sch1998}; \citealt{Zar2002}; \citealt{Isr2010}; \citealt{Has2011}; \citealt{Sub2012},  \citealt{Deb2017}). In this study we have obtained our own estimation of the reddening for the  2997 RR Lyrae variables in our sample from the difference between their intrinsic and observed colours:
\begin{equation}\label{eq:red}
E(V-I)=(V-I)-(V-I)_0
\end{equation}
 \citet{Pier2002} developed an empirical relation that connects the intrinsic colour $(V-I)_0$ of RRab stars to the $V$-band amplitude and the period:

\begin{equation}\label{eq:vi0}
(V-I)_0=(0.65\pm0.02) - (0.07\pm0.01)Amp(V) + (0.36\pm0.06)\log P
\end{equation}

The OGLE~IV catalogue provides amplitudes in the $I$ passband. We transformed  them to $V$-band amplitudes using the relation developed as a byproduct during the work for the {\it Gaia} first data release (DR1, \citealt{Clem2016}). The relation was derived using a large sample of RR Lyrae stars which also includes SMC's RR Lyrae variables (Ripepi et al. in preparation). 

\begin{equation}\label{eq:ampv}
Amp(V)=(1.487\pm0.013)Amp(I) + (0.042\pm0.007)
\end{equation}

The r.m.s. of the relation is 0.03 mag. We calculated the intrinsic colours of the  2997 RR Lyrae variables in our sample using Eqs.~\ref{eq:vi0}-\ref{eq:ampv}. The uncertainties  in $(V-I)_0$ were calculated by error propagation adopting the r.m.s. of Eq.~\ref{eq:ampv} as an uncertainty in $Amp(V)$. Then we calculated individual reddening values for all  2997 variables using Eq.~\ref{eq:red} and $V$ and $I$ apparent magnitudes from OGLE~IV. Uncertainties  in apparent $V$ and $I$ magnitudes are not provided in the OGLE~IV catalogue. Following \citet{Jac2016} we assumed their values  as 0.02~mag.  Reddening values and related uncertainties  obtained by this procedure are provided in Table~\ref{tab:gen}.

 For  27 RRab stars in the sample it was not possible to estimate the reddening since their $V$ apparent magnitudes are not available in the OGLE~IV  catalogue.  The mean reddening of the remaining  2970 RR Lyrae variables is $\langle E(V-I) \rangle =0.06\pm0.06$ mag. We assigned this mean value to the  27 stars for which a direct determination of reddening was not possible. 

 Reddening estimates bear uncertainties which may lead to negative reddening values. For  232 stars out of  2997 (8\% of the sample) we found negative reddening values. However,  165 of them have reddening consistent with zero within the uncertainties, which means that only  67 variables (2\% of the sample) have really negative values.  The lowest reddening in the sample is $E(V-I)=-0.56$~mag, while  the median reddening of all stars with negative values is  $E(V-I)=-0.02$~mag. If negative reddening values are ignored, this will skew the distribution towards 
positive reddening values and lead to biases. By retaining negative reddening values, the uncertainties will be propagated properly. Hence, we keep the negative values as a reflection of the uncertainties in reddening in the following analysis.
%

\begin{figure}
\includegraphics[trim=10 80 30 30 clip, width=\linewidth]{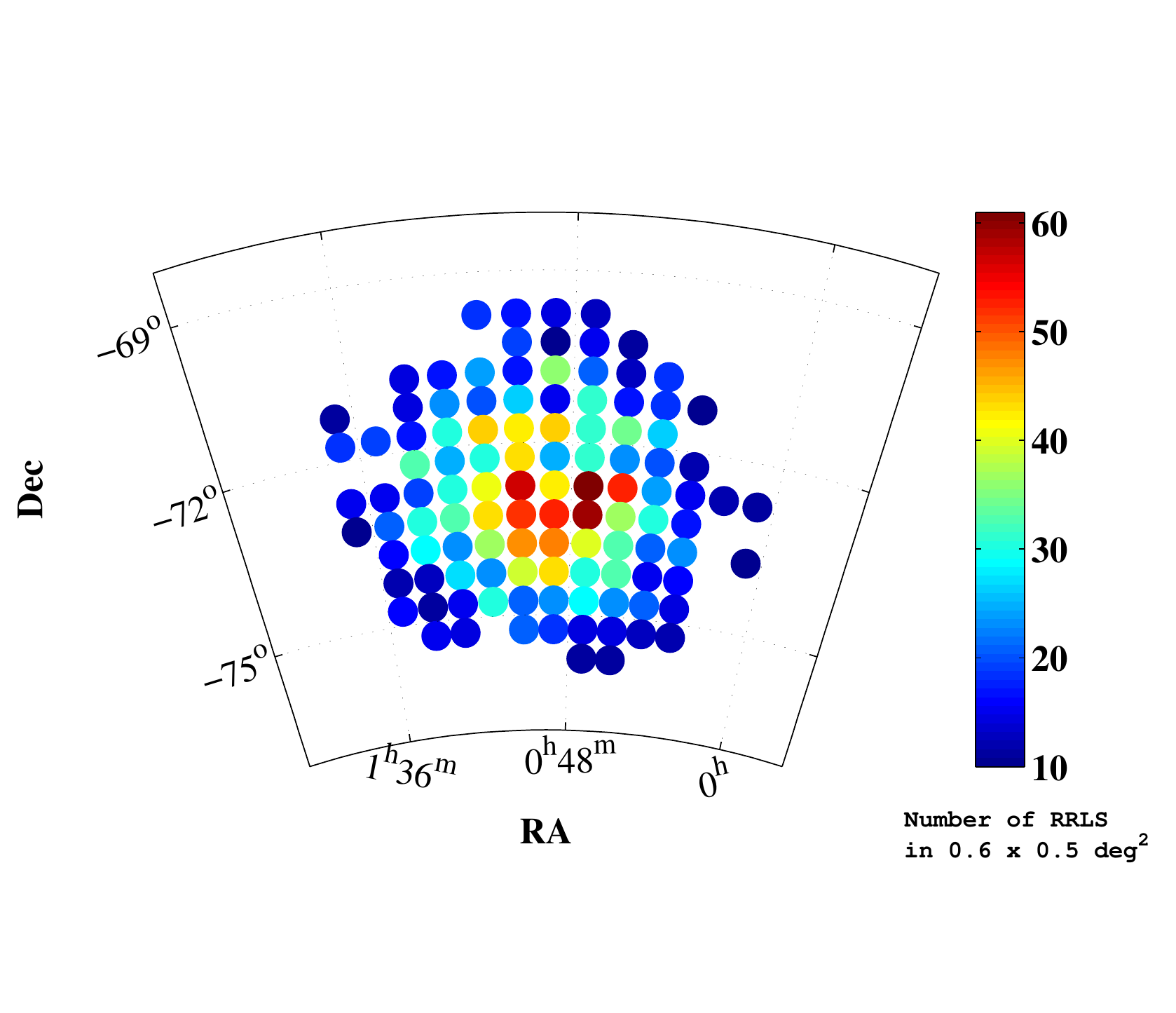}
\caption{ Two-dimensional distribution of the number of stars in different sub-regions of the SMC. Each point corresponds to a sub-region of  0.6 $\times$ 0.5 deg$^2$ area, for a total number of 110 sub-regions.\label{fig:numb}}
\end{figure}

\begin{figure}
\includegraphics[trim=10 80 30 30 clip, width=\linewidth]{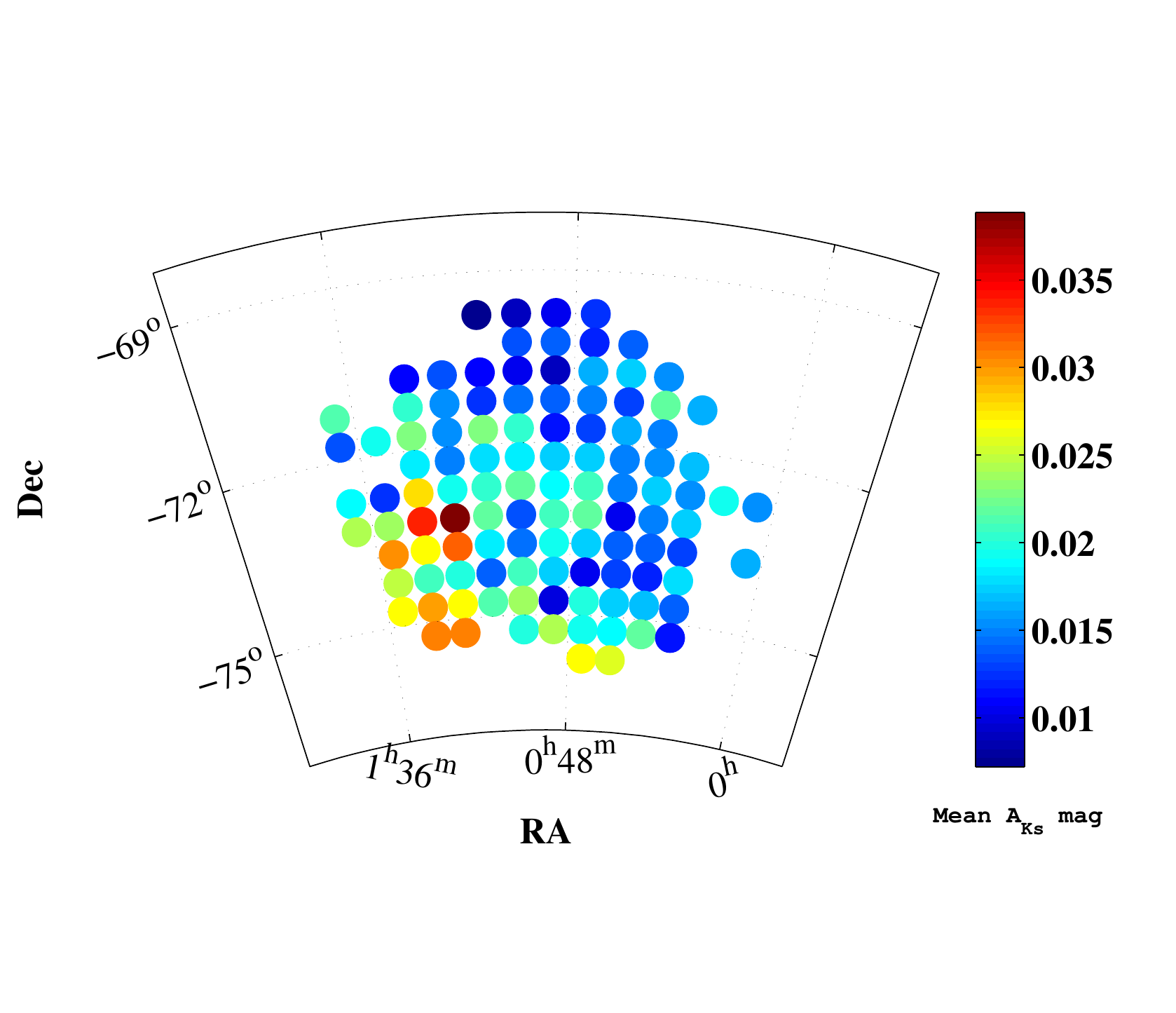}
\caption{Two-dimensional distribution of the mean extinction measured in different sub-regions of the SMC. Each point corresponds to a sub-region of  0.6 $\times$ 0.5 deg$^2$ area, for a total number of  110 sub-regions.\label{fig:mean_ext}}
\end{figure}

 We have divided the SMC region in small sub-regions of equal area (0.6 $\times$ 0.5 deg$^2$). Only those sub-regions which have at least 10 RR Lyrae variables (3 $\times$ Poissonian error) are considered for the analysis. There are 110 sub-regions which satisfy this criterion. The number of stars in sub-regions ranges from 10 to 60. More specifically, the number of stars is 10$-$30 in the outer regions and 30$-$60 in the inner regions of the SMC. Fig.~\ref{fig:numb} shows the distribution of  stars in each of the 110 sub-regions. In Fig.~\ref{fig:mean_ext} we show the mean extinction values of RR Lyrae variables located in  the sub-regions. The extinction is larger in the eastern/south-eastern parts of the SMC. Our extinction maps are very similar to the extinction maps produced by \citet{Sub2012} and \citet{Has2011} using RC stars,  and \citet{Deb2017} using RR Lyrae stars. We discuss this in details in Section~\ref{sec:dist_map}.

In order to calculate dereddened $K_{\rm s,0}$ magnitudes we applied the relations $E(V-I)=1.22E(B-V)$ and $A_{K}=0.114A_{V}$  (\citealt{Card1989}, \citealt{Caputo2000}) thus obtaining:
\begin{center}
\begin{equation}
K_{\rm s,0}=K_{\rm s} - 0.29E(V-I)
\label{eq:ext}
\end{equation} 
\end{center}

The mean extinction $A_{K_{\rm s}}$ of the  2997 RR Lyrae stars in our sample is 0.02~mag, which is significantly smaller  than the typical uncertainty of the $K_{\rm s}$ individual mean apparent magnitudes 
(0.07~mag). 
 \citet{Sch1998} estimated the typical reddening towards the SMC from the median dust emission in surrounding annuli and found the value $E(B-V)=0.037$~mag which corresponds to $E(V-I)=0.045$~mag. This value is smaller than the value of reddening estimated in this paper $E(V-I)=0.06\pm0.06$~mag, but is in agreement with it within the errors. \citet{Zar2002} produced an extinction map across the SMC and studied  the nature of extinction as a function of stellar population. In particular, these authors derived extinction values for cooler and older stars  ($5500~{\rm K} \le T_{eff} \le 6500~{\rm K}$) and for hotter and younger stars ($12000~{\rm K} \le T_{eff} \le 45000~{\rm K}$).
They found that the mean extinction is lower for the cooler population  ($A_V$=0.18~mag). This corresponds to $A_K=0.02$~mag, which is equal to the mean extinction value that we found for our sample of 2997 RR Lyrae stars in the SMC. \citet{Isr2010} estimated the mean internal extinction of the SMC: $A_V=0.45$~mag, which corresponds to $A_K=0.05$~mag, hence, higher than the mean extinction $A_K=0.02$~mag found in this paper for the RR Lyrae stars. \citet{Sub2012} estimated the  reddening towards  the SMC using RC stars and found the mean value  $E(V-I)=0.053\pm0.017$~mag. \citet{Has2011}  found the mean reddening of the SMC to be  $E(V-I)=0.04\pm0.06$~mag using RC stars and  $E(V-I)=0.07\pm0.06$~mag from the RR Lyrae stars. \citet{Deb2017} estimated the mean reddening value of the SMC using OGLE~IV RR Lyrae stars and found $E(B-V)=0.056\pm0.019$~mag, which corresponds to  $E(V-I)=0.068$~mag. All the estimates of the reddening in the SMC based on RC and RR Lyrae stars appear to be consistent with the value found in the present study.

Individual dereddened $K_{\rm s,0}$ magnitudes  of 2997 RR Lyrae stars in the SMC are listed in column 7 of Table~\ref{tab:gen}  and used in the following analysis to derive the individual distance to each RR Lyrae star in the sample. 

\section{Period-luminosity relation}
\label{sec:pl}
 

We performed a  non-weighted linear least squares fit of the $PK_{\rm s,0}$ relation defined by the  2997 SMC RR Lyrae variables in our sample:
\begin{equation}\label{eq:PL_SMC}
 K_{\rm s,0} = (-3.17 \pm 0.08)\textrm{log}P + (17.68 \pm 0.02)
\end{equation}

\begin{figure}
 \includegraphics[trim=50 170 30 100 clip, width=\linewidth]{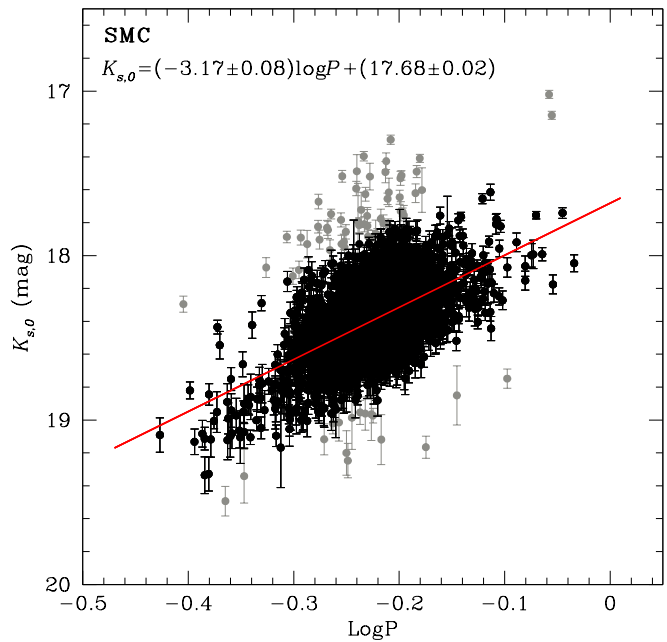}
  \caption{$PK_{\rm s,0}$ relation of  2997 RR Lyrae stars in the SMC. The relation was calculated as a non-weighted linear regression by progressively discarding objects which deviate more than $3\sigma$. Black and grey dots represent objects located within and beyond $3\sigma$ from the regression line, respectively. The red line shows the best fit.}
  \label{fig:PL_all}
\end{figure}

We used the non-weighted fit in order to avoid biasing by brighter objects which usually have smaller uncertainties. 
The fit is shown in Fig.~\ref{fig:PL_all}. The r.m.s.  of the relation is large (0.17 mag) and could be owing to: metallicity differences, intrinsic dispersion of the $PK_{\rm s,0}$ relation or depth effect. 
 The dependence of the $K$-band magnitude on metallicity has been investigated in a number of studies with a tendency of the theoretical and semi-theoretical analyses (\citealt{Bono2003}; \citealt{Cat2004}) to derive a steeper metallicity slope  than found by  the empirical  analyses (\citealt{DelP2006}; \citealt{Sol2006,Sol2008}; \citealt{Bor2009}; \citealt{Mur2015}). Literature empirical values for the metallicity slope vary from $0.03 \pm 0.07$ (\citealt{Mur2015} based on 70 field RR Lyrae variables in the bar of the LMC) to $0.12 \pm 0.04$ (\citealt{DelP2006} from the analysis of RR Lyrae stars in $\omega$ Cen).
Thus, the dependence of the $K_{\rm s}$ magnitude  on metallicity does not seem to be able to explain the large scatter seen in Eq.~\ref{eq:PL_SMC}. 
Hence, we conclude that the large spread observed in the RR Lyrae $PK_{\rm s}$ relation is mainly caused by the intrinsic dispersion of the relation and by depth effect.  A detailed discussion of the SMC line-of-sight depth and the intrinsic dispersion of the $PK_{\rm s,0}$ relation is provided in Section~\ref{sec:los}.

\begin{landscape}

\begin{figure}

\centering 

\includegraphics[trim=220 150 230 160,height=1\textwidth,width=1\textwidth]{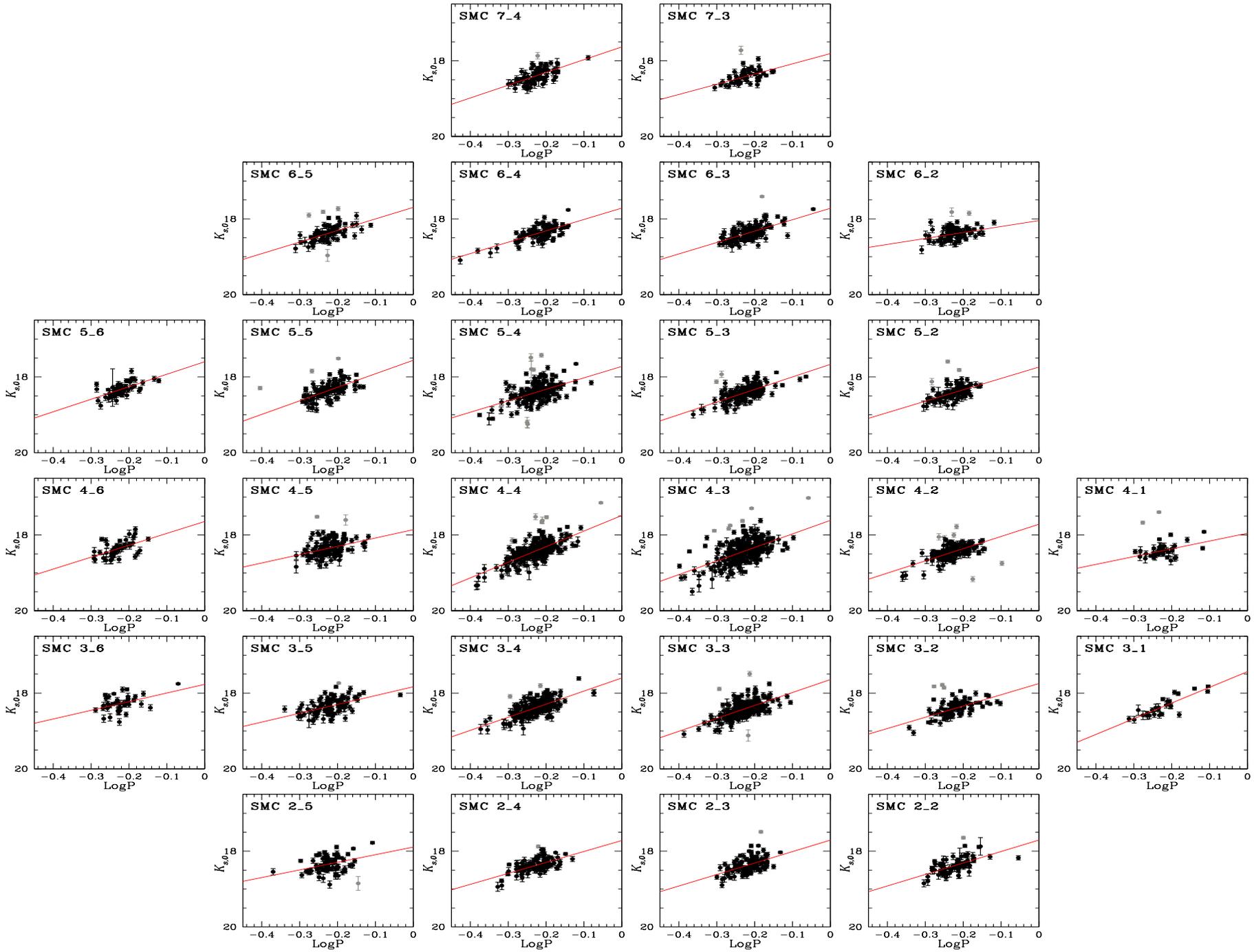}

\caption{$PK_{\rm s,0}$  relations of RR Lyrae stars in each of the 27 VMC tiles of the SMC. The relation for each tile  was calculated as a non-weighted linear regression  by progressively discarding objects which deviate more than $3\sigma$. Black and grey dots represent objects located within and beyond $3\sigma$ from the regression line, respectively. Red lines show the best fits.}
 \label{fig:PL_tiles}

\end{figure}

\end{landscape}

To reduce the scatter we calculated the $PK_{\rm s,0}$ relations as a non-weighted linear regression by progressively discarding objects that deviate more than $3\sigma$,  in each of the  27  SMC tiles selected for the present analysis, separately. The resulting relations are summarised in  column~6 of Table~\ref{tab:tiles} and shown in Fig.~\ref{fig:PL_tiles}. The r.m.s. of the relations remains significant (0.13-0.24 mag) even for single tiles and is systematically larger in the inner regions of the SMC. The slope of the  $PK_{\rm s,0}$ relation in different tiles varies from $-1.57$ to $-4.14$. This provides hints of an elongated structure of the SMC, which we studied using individual distances to the  2997 RR Lyrae stars in our sample. 


\section{Structure of the SMC} 
\label{sec:str}

\subsection{Individual distances of RR Lyrae stars}\label{sec:dist_map}

Using the Fourier parameters of OGLE~IV light curves  \citet{Skowron2016} estimated individual metallicities on the \citet{J95} and \citet{ZW1984}  metallicity scales for  3560 RRab stars in the SMC. 
We found that  2426 out of  2997 RR Lyrae stars in our sample have individual metallicities estimated by \citet{Skowron2016}. 
 We used the metallicity estimates on the Zinn \& West metallicity scale from \citet{Skowron2016} for  2426 RR Lyrae stars in the sample, and to the  571 RR Lyrae stars without an individual metallicity estimate we assigned the mean metallicity of the SMC RRab stars determined by the same authors: $\rm{[Fe/H]_{ZW84}=-1.85\pm0.33}$ dex.  These 571 stars are distributed smoothly in the SMC with a slight concentration in the bar-like feature, where the crowding could have prevented an accurate estimation of metallicity.

We use our dereddened intensity-averaged $K_{\rm s,0}$ magnitudes (Section~\ref{sec:ext}), periods from the OGLE~IV catalogue \citep{Sos2016} and metallicities from \citet{Skowron2016} along with eq.~16 from \citet{Mur2015} to determine individual distance moduli  for each of the  2997 SMC RR Lyrae stars.
The $PM_{K_{\rm s}}Z$ relation in \citet{Mur2015} was obtained on the metallicity scale defined in  \citet{Grat2004}. \citet{Grat2004} pointed out that there is no clear offset between the metallicity scales defined in their paper and \citet{ZW1984} metallicity scale, and considered the 0.06 dex difference found for the three calibrating clusters used in their analysis as a possible offset. Hence, we added 0.06 dex to the metallicity values provided by \citet{Skowron2016}. However, the dependence on metallicity of the $M_{K_{\rm s}}$ magnitudes is small (0.03, \citealt{Mur2015}), hence, a 0.06 dex offset will give only 0.002~mag error to the distance modulus, which is much smaller than typical errors of distance moduli  (0.15~mag). Thus, the uncertainty in the offset between two metallicity scales will not affect our final results.  Individual distance moduli obtained for the 2997 RR Lyrae stars in the sample are summarised in column 11 of  Table~\ref{tab:gen}.


The weighted mean distance modulus of the  2997 RR Lyrae stars in our sample is $(m-M)_0=18.88$~mag, with a standard deviation of 0.20~mag.  This is 0.08 mag shorter than derived by \citet{deGr2015} based on a statistical analysis of the SMC distance estimates available in the literature,  $(m-M)_0=18.96\pm0.02$~mag, but both determinations are entirely consistent with each other within the mutual uncertainties (standard deviations). It is also shorter than the estimates of the SMC's distance modulus based on classical Cepheids observed by VMC  $(m-M)_0=19.01\pm0.05$~mag, $(m-M)_0=19.04\pm0.06$~mag \citep{Ripepi2016} and $(m-M)_0=19.01\pm0.08$~mag \citep{Marconi2017}. Note that the mean distance modulus derived in the current study reflects only the statistical distribution of individual distance moduli in the sample, and is not a distance to the centre of the ellipsoid formed by RR Lyrae stars. The derived mean value is significantly affected by the spatial distribution of RR Lyrae stars and  cannot be considered as a distance to the SMC. The mean distance modulus derived in the current study corresponds to the mean distance $60.0$~kpc. The standard deviation of the mean value is $\sim5$ kpc and reflects the extension of the SMC RR Lyrae star distribution along the line-of sight (see Section~\ref{sec:los}).

\begin{landscape}

\begin{figure}

\centering 

\includegraphics[trim=220 140 230 150,height=1\textwidth,width=1\textwidth]{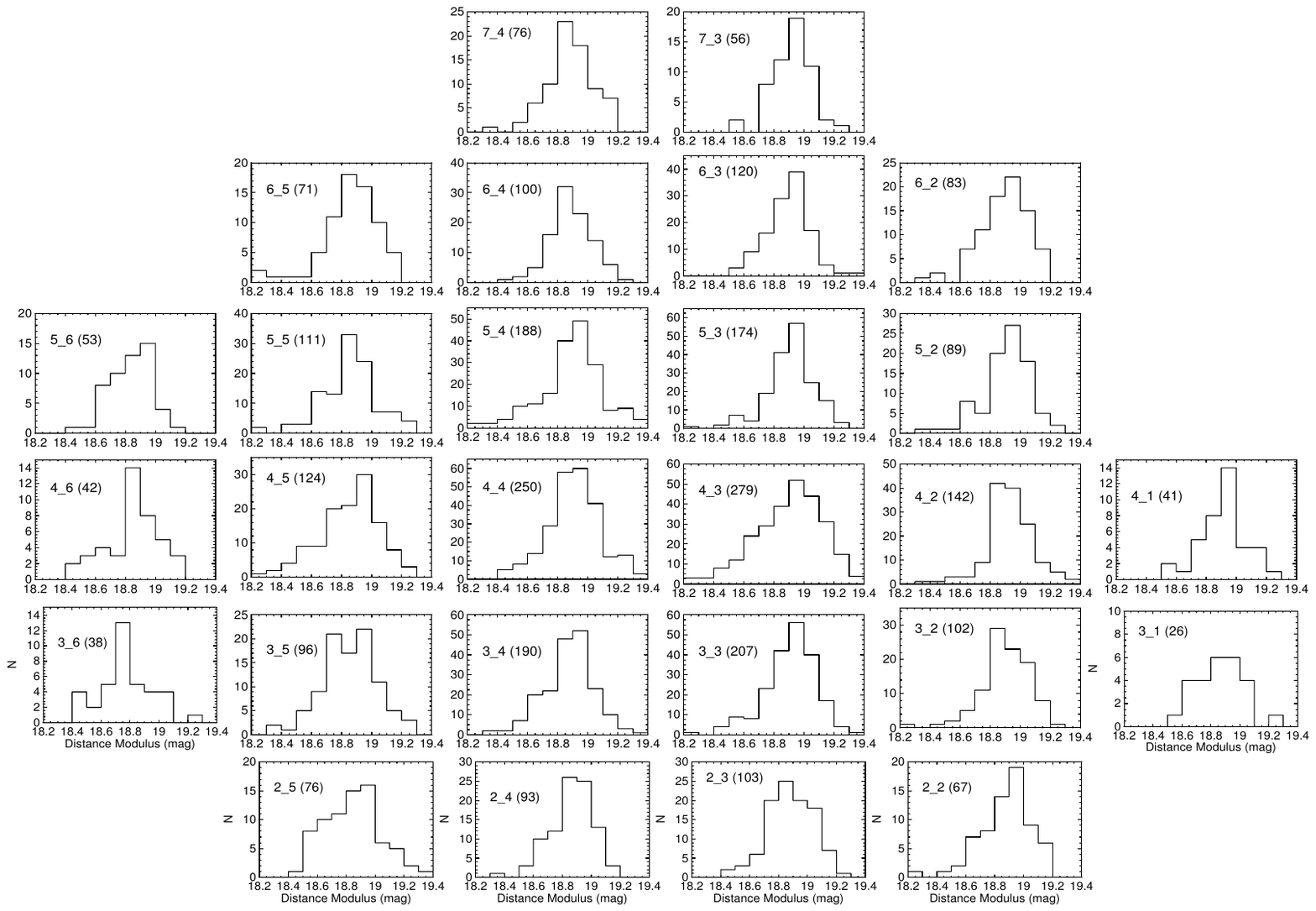}

\caption{Distance modulus distributions of the RR Lyrae stars in the 27 VMC tiles.  The tile number is shown omitting the  SMC  prefix. The total number of RR Lyrae stars  in each tile is shown
in parentheses.\label{fig:dist_tiles}}

\end{figure}

\end{landscape}

We also calculated mean distances to the RR Lyrae stars in each of the  27 tiles, separately (Table~\ref{tab:tiles}). The derived distance moduli span the range from 18.78 to 18.94~mag. 
There is a clear trend of tiles located in the eastern and south-eastern parts of the SMC to be closer to us. However, distances to stars in a given  tile calculated in this way are approximate, since the same distance is assigned to all stars in a tile. To perform a more accurate analysis of the SMC's  structure we used the individual distances of the  2997 RR Lyrae variables. In Fig.~\ref{fig:dist_tiles} we present the distance modulus distributions defined by  the individual RR Lyrae stars in each tile. It shows that the eastern regions have asymmetric distributions of the RR Lyrae stars and are located closer to us.

\begin{figure*}
\includegraphics[width=1.0\textwidth]{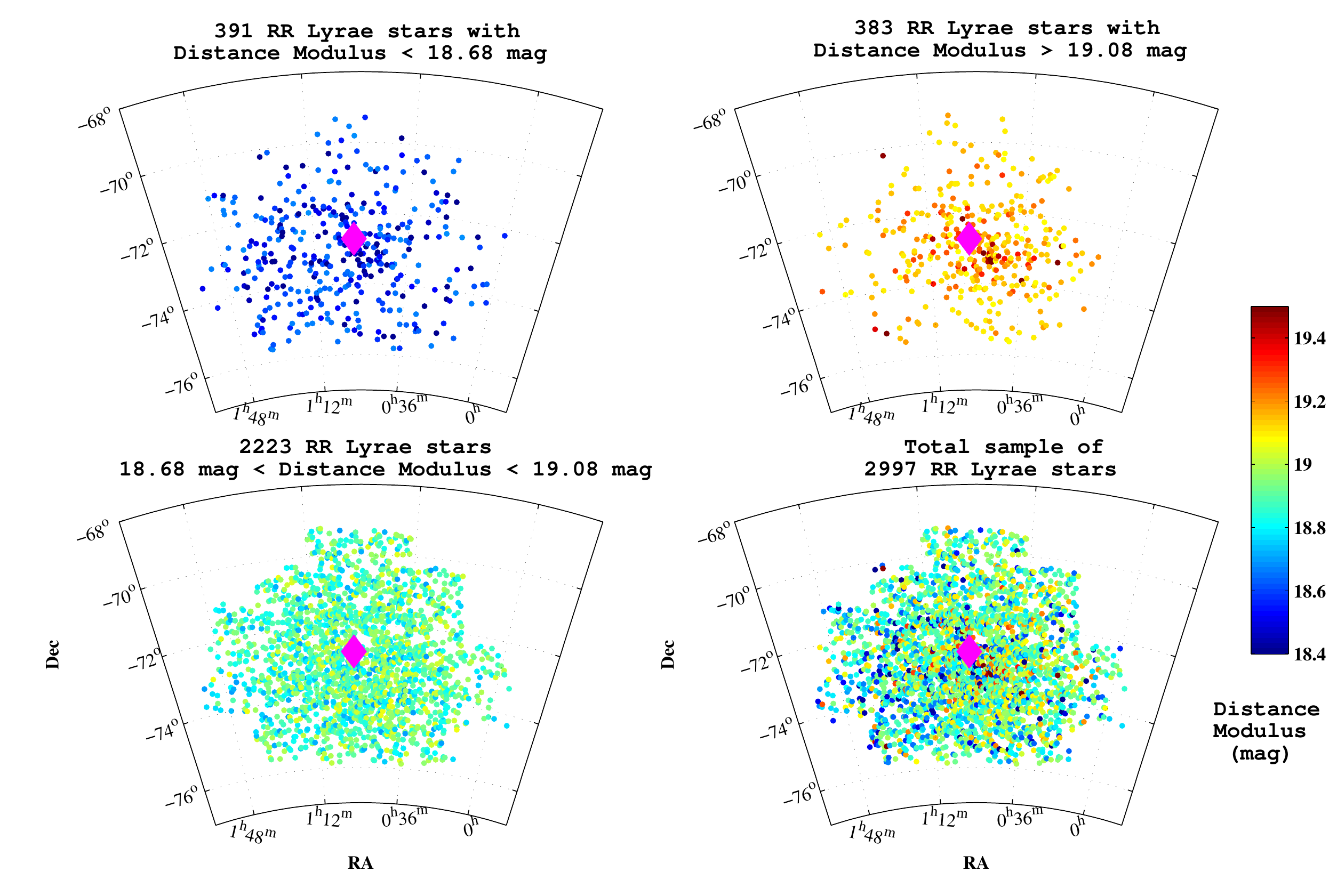}

\caption{Two-dimensional distribution of the RR Lyrae stars' distance moduli. The upper-left, upper-right, lower-left and lower-right panels show respectively the closer RR Lyrae stars with $(m-M)_0<18.68$~mag, the more distant RR Lyrae stars with $(m-M)_0 >19.08$~mag, the sample within 1$\sigma$ error of the mean distance modulus and the total sample. A magenta diamond in all panels represents the centroid of the sample.\label{fig:dist}}
\end{figure*}

The two-dimensional distribution of distances in the SMC is shown in Fig~\ref{fig:dist}. The upper-left, upper-right, lower-left and lower-right panels  show respectively the closer RR Lyrae stars with $(m-M)_0<18.68$~mag, the more distant RR Lyrae stars with $(m-M)_0 >19.08$~mag, the sample within 1$\sigma$ error of the mean distance modulus and the total sample. RR Lyrae stars at all distances are distributed smoothly. We cannot see any signature of the bar from the distribution of RR Lyrae stars.

\begin{figure*}
\includegraphics[width=1.0\textwidth]{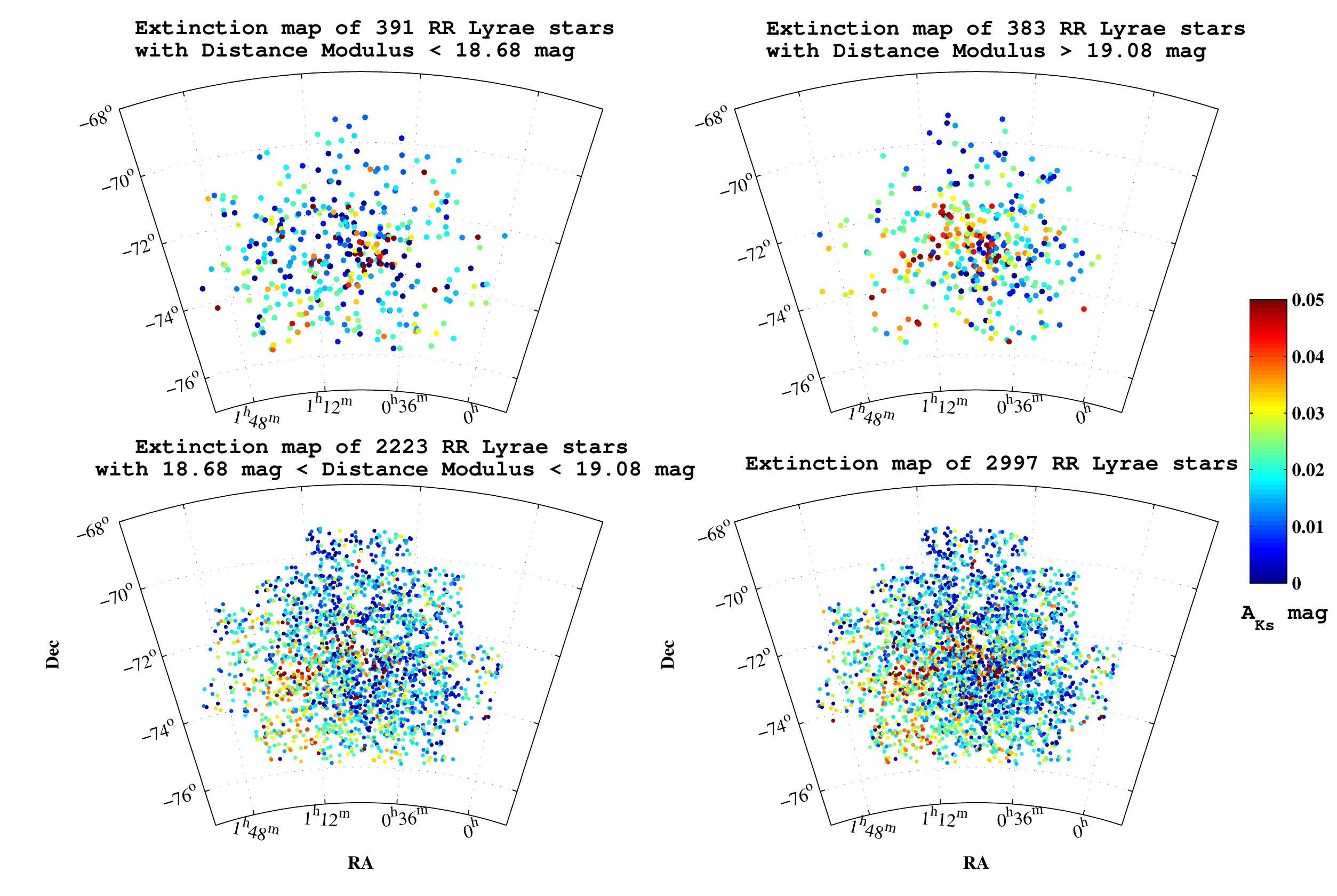}
\caption{Two-dimensional distribution of the RR Lyrae stars' extinction $A_{K_{\rm s}}$. The upper-left, upper-right, lower-left and lower-right panels show respectively the closer RR Lyrae stars with $(m-M)_0<18.68$~mag, the more distant RR Lyrae stars with $(m-M)_0 >19.08$~mag, the sample within 1$\sigma$ error of the mean distance modulus and the total sample.\label{fig:map_ext}}
\end{figure*}

Similarly to Fig.~\ref{fig:dist}, in Fig.~\ref{fig:map_ext} we plot the two-dimensional distribution of extinction, $A_{K_{\rm s}}$, derived in Section~\ref{sec:ext}.  The upper-left and upper-right panels of Fig.~\ref{fig:map_ext} show that more distant RR Lyrae stars in general have higher values of extinction than closer stars. The mean extinction of the closest RR Lyrae variables is $\langle A_{K_{\rm s}} \rangle =0.017$~mag which should be compared with $\langle A_{K_{\rm s}} \rangle =0.023$~mag derived for the most distant stars. The highest extinction is observed in the central part and in the active star-forming region in an eastern extension of the SMC called the Wing, located at $\alpha$ = $01^{\rm h} 15^{\rm m}$ and $\delta$ = $-$73$^{\circ}$10$^\prime$ \citep{Has2011}. The concentration of stars with high extinction in the central regions seen in all panels of Fig.~\ref{fig:map_ext} extends from the north-east to south-west and outlines  the bar-like feature of the SMC. Thus, although we could not detect the bar from the spatial distribution of RR Lyrae stars (Fig.~\ref{fig:dist}), the bar-like feature is seen in the two-dimensional  distribution of the extinction (Fig.~\ref{fig:map_ext}). Note that a higher extinction in the Wing of the SMC  is seen for all RR Lyrae stars except the closest ones, which could mean that the star-forming region is located at a greater distance than $(m-M)_0=18.68$~mag. These findings are in agreement with results obtained by \citet{Has2011} and \citet{Sub2012} who measured the reddening in the SMC  using RC stars detected  by OGLE~III and  found the main concentrations of higher reddening in the bar and Wing of the SMC.  \citet{Deb2017} determined individual reddening for OGLE~IV RR Lyrae stars and found the south-eastern part of the SMC to contain the  regions of the higher extinction, which is in agreement with our findings.

\begin{figure}
\includegraphics[trim=10 30 20 30  clip, width=\linewidth]{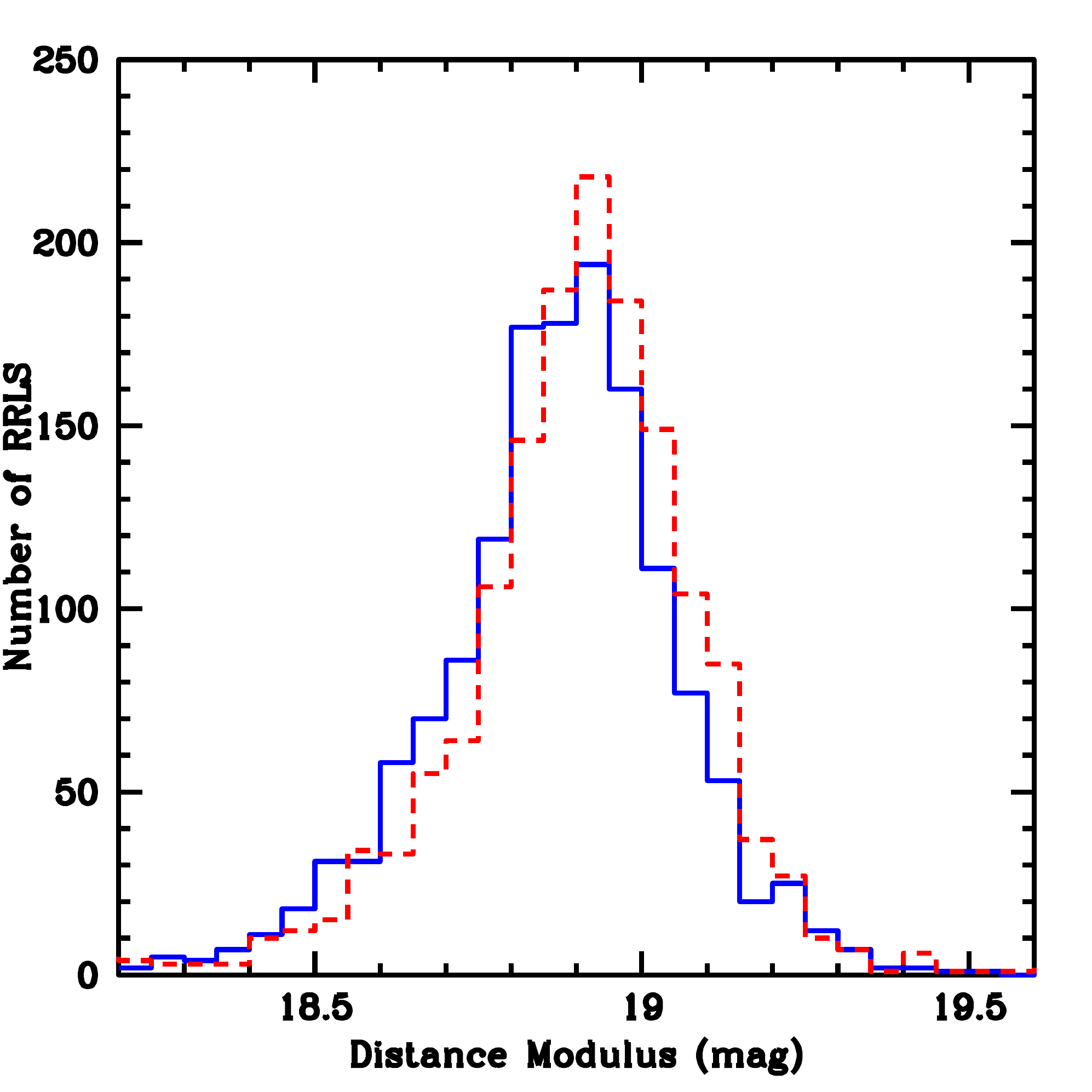}
\caption{Normalised distance distributions of the RR Lyrae stars in the eastern (blue line) and western (red line) regions of the SMC.\label{fig:hist}}
\end{figure}

We calculated the centroid of our sample as an average of RA and Dec values of the RR Lyrae stars, and found that it is located at $\alpha_0$ = $00^{\rm h} 55^{\rm m} 50^{\rm s}.97$ and  $\delta_0$ = $-$72$^{\circ}$51$^\prime$29$\rlap{.}^{\prime\prime}$27. We divided the whole sample of RR Lyrae stars according to RA in an eastern (RA $>$ $\alpha_0$) and a  western (RA $<$ $\alpha_0$) region based on their positions. There are  1473 and 1524 RR Lyrae stars in the eastern and the western regions, respectively. The distance modulus distributions (for a bin size of 0.05 mag) of the RR Lyrae variables in the eastern (blue line) and western (red dotted line) regions of the SMC are shown in Fig.~\ref{fig:hist}. The two distributions show a difference with an excess (15\%) of closer RR Lyrae variables in the eastern region.  We note that the error in individual distances has a range from 0.1 to 0.2 mag, with an average error of 0.15~mag. This is larger than the bin size used in Fig.~\ref{fig:hist}. We changed the bin size from 0.05 to 0.2 mag and we do observe an excess of $\sim$ 15\% closer RR Lyrae stars in the eastern regions for any choice of the bin size.

The mean distances estimated to the eastern and  western regions of the SMC are, respectively,   59.56 $\pm$ 0.14 kpc and 60.50 $\pm$ 0.14 kpc, where errors are calculated as the standard deviation divided by the square root of number of stars in the region. 
This suggests that the eastern region is  $\sim$ 0.94 $\pm$ 0.20 kpc closer to us than the western region. 
We considered whether this asymmetry could be caused by an overestimation of the extinction in the eastern region. In order to check this possibility we assigned the mean value of extinction  derived in Section~\ref{sec:ext} $\langle A_{K_{\rm s}} \rangle $=0.02~mag to all  2997 stars in the sample and derived mean distances to the eastern and western regions of the SMC of $59.57\pm0.14$~kpc and $60.40\pm0.14$~kpc, respectively. These distances are in agreement with the values obtained by applying individual extinction corrections and proves that our distance determination from the $K_{\rm s}$ magnitudes is not affected by a variation of extinction in the SMC. Furthermore, there is still a difference of $0.83\pm0.20$~kpc between the mean distances of RR Lyrae stars in the eastern and western regions. Thus, the appearance of the eastern region closer to us is not owing to an overestimation of the extinction and rather reflects the actual spatial distribution of the SMC RR Lyrae stars. This is consistent with results obtained by \citet{Sub2012}, \citet{Has2012} and  \citet{Deb2015} from the study of SMC RR Lyrae stars  using the OGLE~III data. Recent results from \citet{Jac2016} based on RR Lyrae stars using the OGLE~IV catalogue 
do not imply any sub-structures  but show some asymmetry in the equal density contours in the eastern part of the  SMC.  They detect a shift in the centre of the ellipsoidal fit, to RR Lyrae stars in different density bins, towards the east and closer to the observer as a function of radius from the SMC centre. This again suggests that the eastern region of the SMC includes a larger  number of closer RR Lyrae stars in the entire OGLE IV data set, in agreement with our results. 

\subsection{Mean distances and line-of-sight depth\label{sec:los}}

The variation of the line-of-sight depth across the SMC can be obtained by measuring the mean distance and standard deviation in smaller sub-regions. We divided the SMC region covered by the 27 VMC tiles into smaller sub-regions of equal area (0.6 $\times$ 0.5 deg$^2$), as anticipated in Section~\ref{sec:ext} (110 sub-regions in total). 
 The standard deviation with respect to the mean distance is a measure of the line-of-sight depth. Along with the actual line-of-sight depth in the distribution, the standard deviation has contributions from the intrinsic magnitude spread of RR Lyrae stars owing to errors in photometry and metallicity effect, as well as intrinsic luminosity variations owing to evolutionary effects. 


To perform an approximate estimation of the intrinsic magnitude spread we analysed the SMC globular cluster NGC~121. This is the oldest SMC cluster. It contains four RR Lyrae stars \citep{Walker1988},  which were confirmed to be cluster's members based on their location close to the cluster's centre and their distribution on the Horizontal Branch of the cluster's CMD (Clementini, private communication).  As part of our analysis we derived dereddened $K_{\rm s,0}$ magnitudes for three of them: V32, V35 and V37 (identification from \citealt{Walker1988}). The RR Lyrae star  V36 was not included in our analysis since it turned out to be an RRc star according to the OGLE~IV catalogue \citep{Sos2016}. We used  $K_{\rm s,0}$ magnitudes from VMC, periods from the OGLE~IV catalogue, the metallicity value of  [Fe/H]=$-$1.51~dex \citep{ZW1984} and eq.~16 from \citet{Mur2015} to determine individual distance moduli of the RR Lyrae stars in NGC~121. The 
mean distance modulus of NGC~121 from the RR Lyrae stars is $(m-M)_0=18.97\pm0.07$~mag. Since the three variables belong to the same cluster we consider the depth effect negligible, hence, the standard deviation of the mean value,  0.07~mag,  represents the intrinsic magnitude spread caused by the effects described above. 
\citet{Clementini2003} estimated the intrinsic  $V$-magnitude spread of 101 RR Lyrae stars in the LMC caused by the internal photometric errors, metallicity distribution and evolutionary effect as 0.10~mag. Since the dependence of the $K_{\rm s}$ magnitudes on metallicity and evolutionary effect is smaller than in $V$ band, our estimation of intrinsic dispersion in the SMC as 0.07~mag is reasonable, even though it is based only on three stars.

\begin{figure}
\includegraphics[trim=10 70 30 30 clip, width=\linewidth]{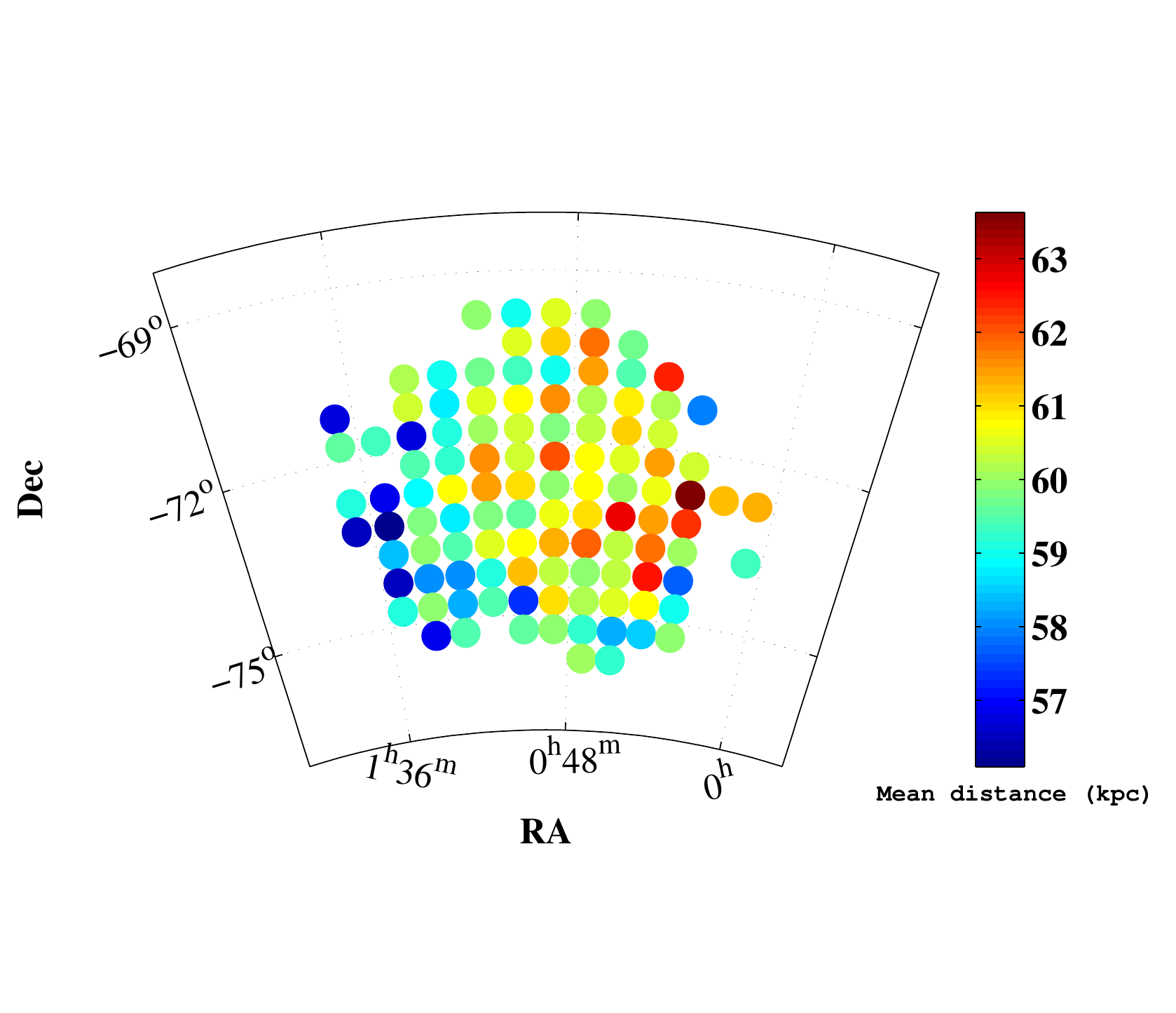}
\caption{Two-dimensional distribution of the mean distance measured in different sub-regions of the SMC. Each point corresponds to a sub-region of  0.6 $\times$ 0.5 deg$^2$ area, for a total number of  110 sub-regions. \label{fig:distr_dist}}
\end{figure}
\begin{figure}
\includegraphics[trim=10 70 30 20 clip, width=\linewidth]{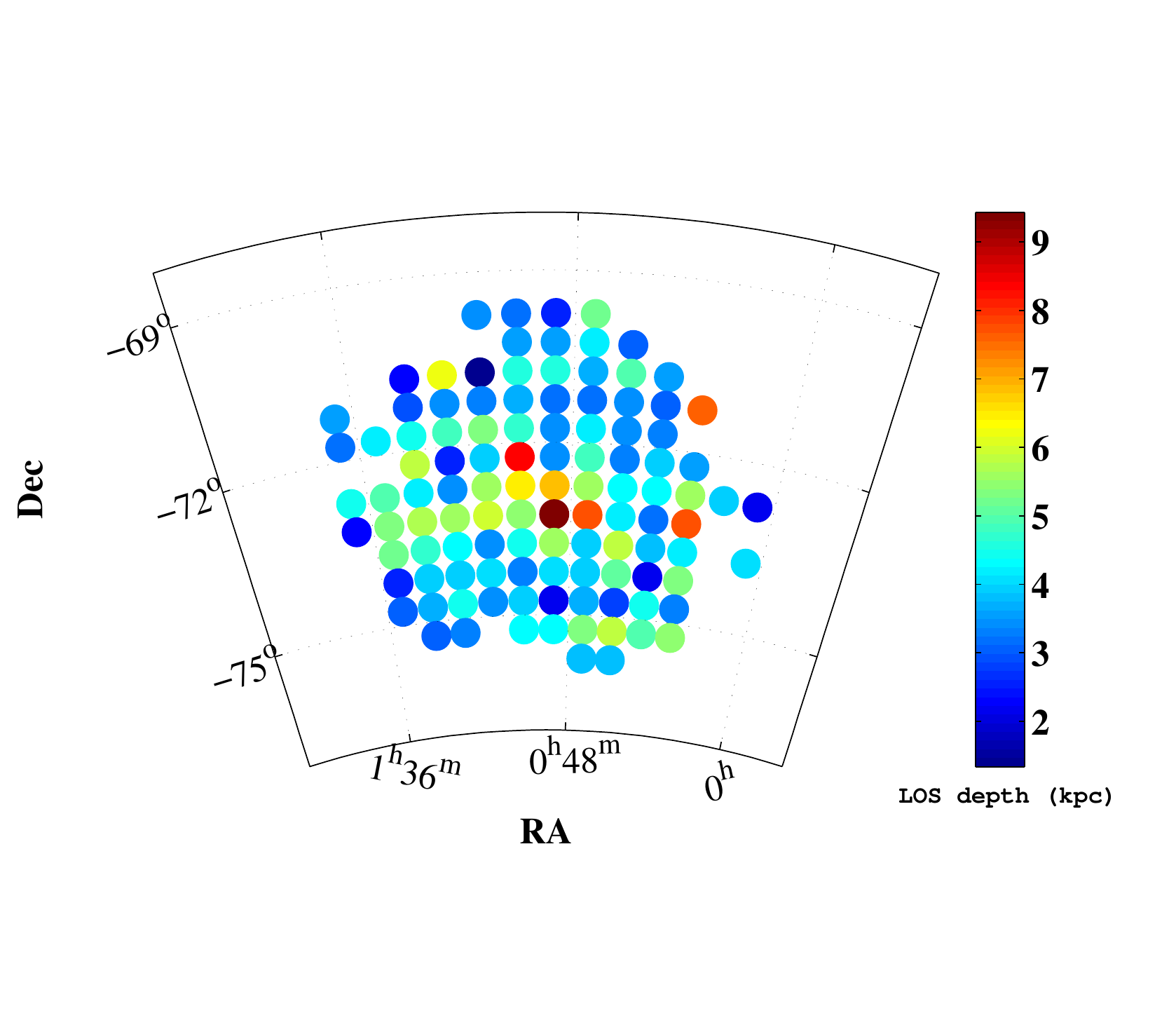}
\caption{Two-dimensional distribution of the line-of-sight depth measured in different sub-regions of the SMC. Each point corresponds to a sub-region of 0.6 $\times$ 0.5 deg$^2$ area, for a total number of  110 sub-regions.\label{fig:distr_los}}
\end{figure}

 The actual line-of-sight depth in the 110  SMC sub-regions is estimated as $\sigma^{2}_{\rm los} = \sigma^{2}_{\rm measured} - \sigma^2_{\rm intrinsic}$. Figs.~\ref{fig:distr_dist} and \ref{fig:distr_los} show the two-dimensional distributions we obtain for  mean distance and actual line-of-sight depth, respectively. The mean distance map clearly shows that the eastern sub-regions are closer to us, as it was widely discussed above. The actual line-of-sight depth values range from  1 to 10~kpc, with an average depth of  4.3 $\pm$ 1.0 kpc. Fig.~\ref{fig:distr_los}  shows that the largest depth is found in the central regions of the SMC. The standard deviation associated with the mean distance modulus of our entire sample is 0.2~mag (Section~\ref{sec:dist_map}). After correcting for the intrinsic width, this dispersion corresponds to a line-of-sight depth of 5.2~kpc. 
 
 The tidal radius of the SMC is estimated as 4 -- 9 kpc by \citet{Stan2004}, 7 -- 12 kpc by \citet{Sub2012} and \citet{DeP2010} estimated the SMC edge in the eastern direction to be at a radius of 6 kpc. \citet{Nid2011} found the presence of old/intermediate-age populations at least up to a radius of 9 kpc.  All these values are comparable  to the 1$\sigma$ depth of 10 kpc observed in the central regions of the SMC.

\begin{figure}
  \includegraphics[trim=30 180 30 100,width=1.05\linewidth]{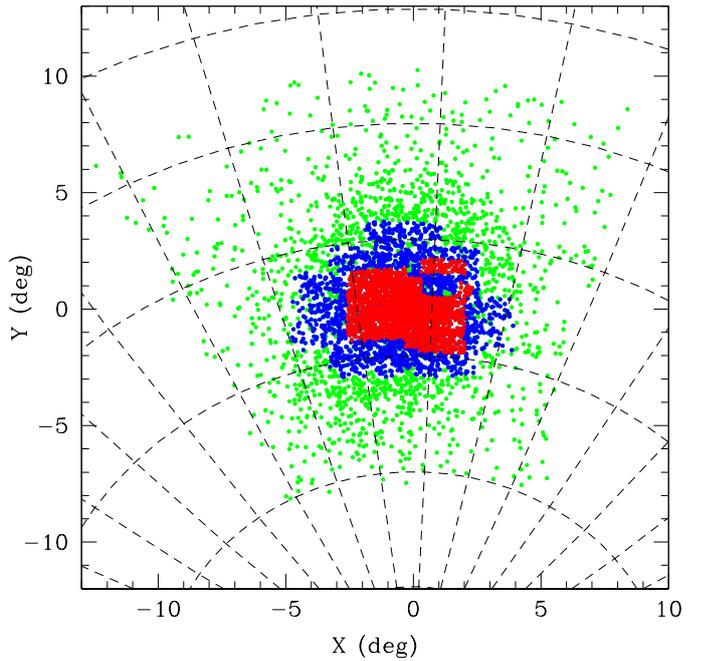}
  \caption{Distribution of RRab stars from the OGLE~IV (green dots) and OGLE~III (red dots) catalogues, and RRab stars used in the current analysis (blue dots). Coordinates are defined as in \citet{vandM2001} where $\alpha_0$ = 12.5 deg, $\delta_0$ = $-$73 deg.}
  \label{fig:map2}
\end{figure}

\begin{figure}
\includegraphics[trim=10 30 20 30 clip, width=\linewidth]{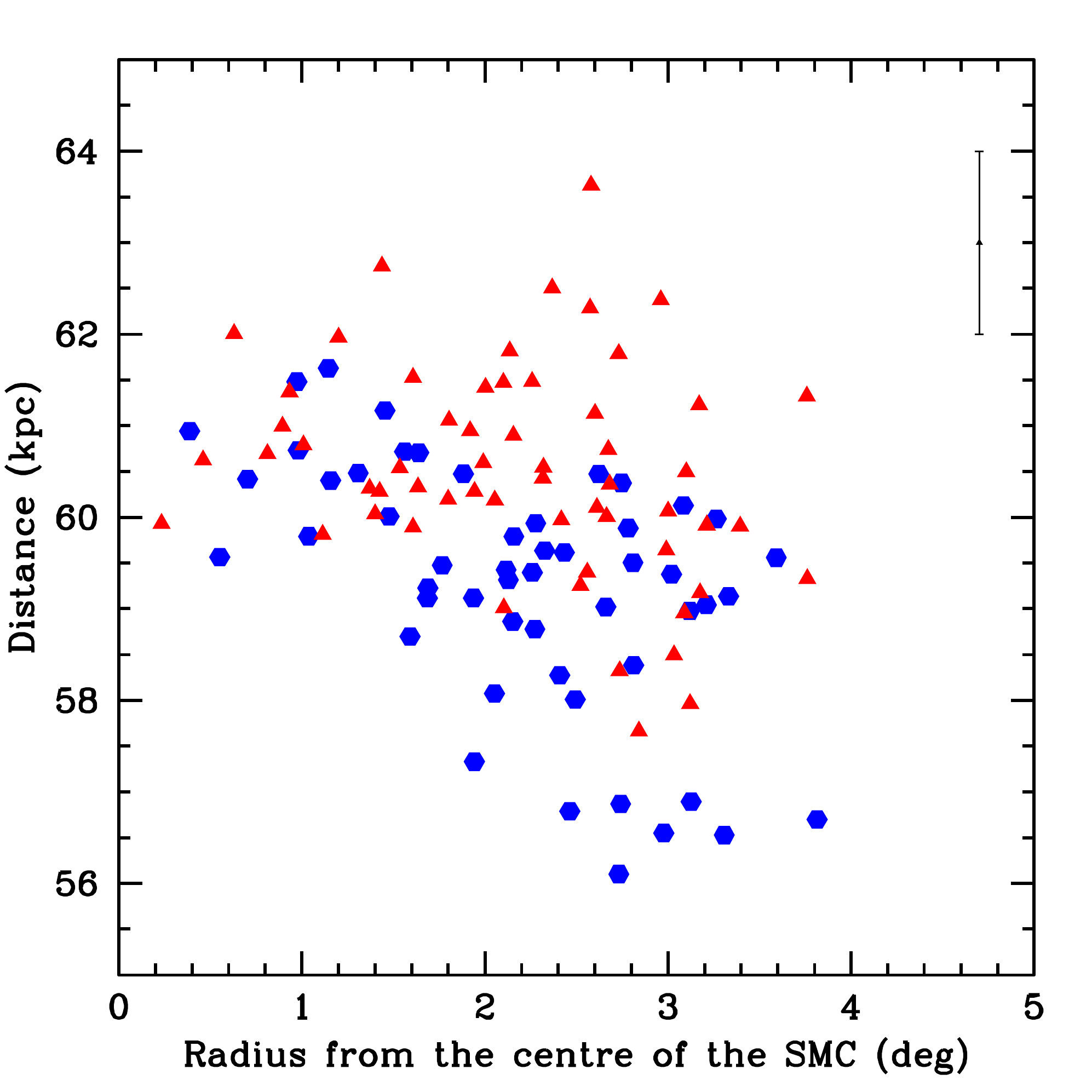}
\caption{Radial variation of the mean distance of the sub-regions. Blue filled hexagons represent regions in the eastern and red filled triangles represent regions in the western parts, respectively. The typical error bar associated with the mean distance is shown in black on the top-right of the panel.\label{fig:grad}}
\end{figure}

\subsection{Three dimensional structure of the SMC}

\begin{table*}
	\centering
\caption{Structural parameters of the SMC ellipsoid as inferred from the RR Lyrae stars\label{tab:par}.}
\label{tab:par}
\begin{tabular}{@{}lcccc@{}}
\hline 
Relation & Axes ratio & $\phi$ & $i$ & Data \\
\hline 
\citet{Sub2012} & $1:1.33:1.61$ & 70$^\circ$.2 & 2$^\circ.$6 & OGLE~III \\
\citet{Has2012} & $-$ &  83$^\circ$$\pm21^\circ$ & 7$^\circ$$\pm15^\circ$  & OGLE~III \\
\citet{Kap2012} & $1:1.23:1.80$ & $-$ & $-$  & OGLE~III \\
\citet{Deb2015} & $1:1.310(\pm0.029):8.269(\pm0.934)$ & 74$^\circ.307\pm0^\circ.509$ & 2$^\circ$$.265\pm0^\circ.784$ & OGLE~III \\
\citet{Jac2016}  & $1:1.10:2.13 $& $-5^\circ - +41^\circ$ & $3^\circ-9^\circ$ & OGLE~IV \\
\citet{Deb2017} & $1.000(\pm0.001):1.100(\pm0.001):2.600(\pm0.015)$ & $38^\circ.500\pm0^\circ.300$ & 2$^\circ$$.300\pm0^\circ.140$ & OGLE~IV\\
\hline
This study & 1:1.11($\pm$0.01):3.30($\pm$0.70) & 84$^\circ$.6$\pm$0$^\circ$.2  & 2$^\circ$$.1\pm$0$^\circ$.6 & OGLE~IV \& VMC\\

\hline
\end{tabular}
\end{table*}

The Cartesian coordinates corresponding to each RR Lyrae star can be obtained using the star's RA, Dec coordinates and the distance modulus. We assume the $x$-axis antiparallel to the RA axis, the $y$-axis parallel to the Dec axis, and the $z$-axis along the line-of-sight with values increasing towards the observer. The distance modulus is used to obtain the distance to each star in kpc. The RA, Dec and the distance are converted into $x$, $y$, $z$ Cartesian coordinates using the transformation 
equations given by \citet{vandM2001} and assuming the origin of the system at $\alpha_0$ = $00^{\rm h} 55^{\rm m} 50^{\rm s}.97$ and  $\delta_0$ = $-$72$^{\circ}$51$^\prime$29$\rlap{.}^{\prime\prime}$27. The distance to the origin of the SMC is taken as 60~kpc with standard deviation of 5~kpc (Section~\ref{sec:dist_map}).  RA and Dec of each star correspond to the VMC  coordinates. The positional accuracy of the VMC survey is $\sim$ 0.02$^{\prime\prime}$. The errors associated with $x$, $y$ and $z$ are calculated using error propagation.  

Based on both observations and theoretical studies (\citealt{Sub2012} and references therein), the old and intermediate-age stellar populations in the SMC are suggested to be distributed in a spheroidal/ellipsoidal system. Our sample of SMC RR Lyrae stars shows a smooth distribution on the sky. Thus we modelled the RR Lyrae stellar distribution as a triaxial ellipsoid. The parameters of the ellipsoid, such as the axes ratio, the position angle of the major axis of the ellipsoid projection  on the sky ($\phi$) and the inclination of the longest axis with respect to the sky plane ($\it{i}$) are estimated using a method of inertia tensor analysis similar to that described in  \cite{Sub2012} (also refer to \citealt{Pejcha2009}; \citealt{Paz2006}). The basic principle is to create an inertia tensor of the $x$, $y$, $z$ coordinates and estimate the Eigen vectors and Eigen values. The Eigen vectors correspond to the spatial directions and the square roots of the Eigen values correspond to the axes ratio.

First, we applied the method to the ($x$,$y$) coordinates and found that the stars are located in an elongated distribution with an axes ratio of  1:1.12($\pm$0.01) and the major axis has a position angle  $\phi$ = 85$^\circ$.5$\pm$0$^\circ$.1. We repeated the procedure with ($x$,$y$,$z$) coordinates. The $z$ values have a large scatter, owing to large uncertainty in distances, compared to good positional accuracy (0.02$^{\prime\prime}$). As described in \cite{Pejcha2009}, to correct for this internal scatter in the $z$ value, we subtracted the internal scatter of 0.07 mag from the z-component (see Section~\ref{sec:los}). 
We obtained axes ratio, $\phi$ and $\it{i}$ values of  1:1.11($\pm$0.01):3.30($\pm$0.70), 84$^\circ$.6$\pm$0$^\circ$.2  and 2$^\circ$$.1\pm$0$^\circ$.6,  respectively. The small inclination angle of the longest axis with respect to the sky plane suggests that the longest axis of the ellipsoidal distribution of the RR Lyrae stars is oriented nearly along the line-of-sight. 

A compilation of structural parameters of the SMC ellipsoid from the RR Lyrae stars available in the literature is provided in Table~\ref{tab:par}. In the majority of these studies the RR Lyrae stars from the OGLE~III catalogue were used, while  \citet{Jac2016} analysed RRab stars in the OGLE IV catalogue, which covers almost entirely the SMC. In the current study we used RR Lyrae variables observed by OGLE~IV which also have counterparts in the VMC catalogue. Qualitatively all previous results indicate that the RR Lyrae stars in the SMC are distributed in an ellipsoid with the longest axis oriented almost along the line-of-sight. Our results also support these findings, but quantitatively there are some differences. This could be mainly owing to the difference in the spatial coverage of the data used in the various studies.  Fig.~\ref{fig:map2} shows the distribution of RRab stars used in the this paper and RRab stars from the OGLE~III and OGLE~IV catalogues. OGLE~IV covers a larger area than VMC and OGLE~III, which are basically focused on the central regions of the SMC, where the majority of RR Lyrae stars are concentrated.

\vspace{0.3cm}
\subsection{Effect of interaction of the Magellanic Clouds}

Mutual interactions between the Magellanic Clouds and the resultant tidal stripping of
material from the SMC are believed to be the most probable scenario for the
formation of the MB (\citealt{Diaz2012}; \citealt{Besla2012}). One of the
challenges to this scenario is the lack of conclusive evidence for the
presence of tidally stripped intermediate-age/old stars in the MB.
 \cite{Nid2013} identified a closer (distance, D $\sim$ 55 kpc) stellar
structure in front of the main body of the eastern SMC, which is located
4$^\circ$ from the SMC centre. These authors suggested that it is the
tidally stripped stellar counterpart of the H~I in the MB. \citet{Sub2017}
also identified a foreground population (at $\sim12\pm2$~kpc in front of the
main body of the SMC), whose most likely explanation is tidal stripping
from the SMC. Moreover, they identified the inner region (at $\sim$ $2-2.5$ kpc
from the centre) from where the signatures of interactions start becoming
evident, thus, supporting the hypothesis that the MB was formed from tidally
stripped material from the SMC. Both these studies were based on RC
stars. Recently, \citet{Ripepi2017} found that the young ($\sim120$~Myr) and old  ($\sim220$~Myr) Cepheids
in the SMC have different geometric distributions. They also found closer
Cepheids in the eastern regions which are off-centred in the direction of
the LMC. They suggest that these results are owing to the tidal interaction
between the Magellanic Clouds.

The RR Lyrae stars which are older (age $\ge$ 10 Gyr) than RC stars (age $\sim$
2 -- 9 Gyr) and Cepheids (100 -- 300 Myr) are also expected to be affected
by tidal interactions. The RR Lyrae stars are well identified and we have
accurate distances to each individual star. The presence of tidal
signatures in the oldest populations would provide strong constraints to
theoretical models which explain the formation of the MS. Hence, we explore
our data set in more detail.

In the present study we see a shift in the mean distance of the RR Lyrae stars in the eastern 
region compared to the western region. We also see that RR Lyrae stars  in the eastern tiles have 
asymmetric distance distributions (Fig.~\ref{fig:dist_tiles}) and  that the mean
distances of the eastern sub-regions are shorter (Fig.~\ref{fig:distr_dist}). Fig.~\ref{fig:grad}
presents the radial variation of the mean distance to the various
sub-regions. Blue filled hexagons represent  sub-regions in the eastern and
red filled triangles represent sub-regions in the western parts of the SMC.
The plot shows that the eastern sub-regions are located at shorter distances compared
to the western sub-regions. Beyond 2$^\circ$ in radius from the centre,
the majority of the eastern sub-regions have shorter distances. This angular
radius corresponds to a linear radius of $\sim$ 2.2 kpc from the SMC centre
and is similar to the radius from where the signatures of interactions, in
the form of bimodality in the distribution of RC stars, start to become
evident \citep{Sub2017}. We point out that the typical error associated
with the mean distances of the sub-regions is $\sim$ 1 kpc (as shown in the
top-right of the plot). Thus, our results indicate that the oldest stellar
population in the eastern part of the SMC, in the direction of the
MB, is affected by the interactions of the Magellanic
Clouds, 200-300 Myr ago,  which is believed to be mainly responsible for
the formation of the MB.

However, in our present study using RR Lyrae stars, we do not see a clear
bimodality found by \citet{Sub2017}  in the RC distribution, where two peaks separated by $\sim$ 12 $\pm$ 2 kpc are observed  in the
eastern part of the SMC. While comparing our Fig.~\ref{fig:dist_tiles} with fig.~6 of
\cite{Sub2017}, the discrepancy is very evident for tiles 6\_5 and 5\_6. This
implies that RC stars and RR Lyrae stars are distributed in different
structural components in the eastern part of the SMC and/or a different origin of the
foreground RC population. \cite{Sub2017} results on the RC stars are based on only 13 VMC
tiles covering the entire central part and some outer regions. We plan to
address the discrepancy observed between RC and RR Lyrae stars distributions
in the future based on  the whole sample of SMC tiles observed by VMC.

Recently, \citet{Bel2017} identified tidal tails around the LMC and the SMC
using {\it Gaia} DR1 \citep{Gaia1, Gaia2}. They found that the SMC's outer
stellar density contours show a characteristic S-shape which is a typical
signature of tidal stripping seen in satellite galaxies. They also
identified a trailing arm (motion of the SMC with respect to the LMC) from
the SMC which extends towards the LMC. This stellar tidal tail contains
 candidate RR Lyrae stars off-set by $\sim$ 5~deg from the gaseous
MB. Both the old and young MB components originate in
the south-eastern part of the SMC ($\sim$ 6~deg east and 3~deg
south from the SMC centre), where the authors claim see the trailing
tail. Though our observed tiles do not cover the region where
\citet{Bel2017} identified the SMC's trailing arm, the eastern and
south-eastern regions where we see closer RR Lyrae stars are in the
direction of the newly identified SMC's trailing arm.


The study of different stellar populations including RR Lyrae stars in the
outer VMC tiles will be useful to confirm and eventually understand this newly identified
tidal signature. According to the simulations, the line-of-sight velocities
of the stellar debris from the SMC and the LMC are significantly different, hence, a detailed spectroscopic study of RR Lyrae stars and other
distance indicators in the eastern regions of the SMC will also provide valuable
information to understand the interaction history of the MS.

\vspace{-5mm}
\section{Conclusions}\label{sec:sum}

In this study we analysed the structure of the SMC for the first time using multi-epoch  near-infrared photometry of RR Lyrae stars observed by  the VMC survey. 
For this analysis  2997 RRab stars distributed in  27 VMC tiles with visual photometry and pulsation periods from the OGLE~IV survey, were used.  The well-sampled light curves in the $K_{\rm s}$ band obtained by VMC allowed us to derive accurate  intensity-averaged magnitudes for each RR Lyrae star in the sample. Individual reddening values  were calculated using the $V$ and $I$ magnitudes from the OGLE~IV survey. We fit the $PK_{\rm s}$ relations of RR Lyrae stars in each tile, separately,  and found a significant dispersion which is mostly owing to a depth effect. 

To study the three-dimensional structure of the SMC we derived individual distances to the 2997 RR Lyrae stars in the sample by applying the near-infrared $PM_{K_{\rm s}}Z$ relation defined in \citet{Mur2015} to  intensity-averaged $K_{\rm s}$  magnitudes from the VMC survey, periods from the OGLE~IV catalogue and photometrically determined metallicities from \citet{Skowron2016} which we transformed  to the metallicity scale defined by  \citet{Grat2004}.  
The RR Lyrae variables in the SMC are found to  have a roughly spheroidal or ellipsoidal distribution. We modelled the distribution of the SMC RR Lyrae stars as a triaxial ellipsoid and found the values of the axes ratio, position angle of the major axis of the ellipsoid projected on the sky  and  inclination of the longest axis with respect to the sky plane  using  inertia tensor analysis. The parameters of the SMC structure are: axes ratio =  1:1.11($\pm$0.01):3.30($\pm$0.70), 84$^\circ$.6$\pm$0$^\circ$.2  and 2$^\circ$$.1\pm$0$^\circ$.6.
The results obtained in this paper  are generally consistent with previous studies. Existing discrepancies are mostly owing  to differences in the area  of the SMC covered by the different studies. 

The actual line-of-sight depth of the SMC has values in the range from  1 to 10~kpc, with an average depth of $4.3\pm1.0$ kpc. The central parts of the SMC have larger depth. 
Taking into account the standard deviation associated with the mean distance modulus of our entire sample  $(m-M)_0=18.88\pm0.20$ mag we estimated a  line-of-sight depth of the SMC as 5.2~kpc.

The spatial distribution of the RR Lyrae stars in our sample does not show features typical of  young stellar populations, such as a bar. However, from the two-dimensional distribution of the extinction at different distances we can see the bar-like feature and the star forming region in the eastern Wing of the SMC. From our analysis of the SMC structure we concluded that the eastern part of the SMC is located closer to us than the western part. In eastern regions, beyond 2~deg in radius
from the centre, the majority of the sub-regions have shorter distances. The regions where we see a  large number of closer RR Lyrae stars are in the direction of SMC's trailing arm newly identified with {\it Gaia} \citep{Bel2017}.
 All these results  indicate that the oldest stellar population in the eastern part of the SMC is affected by interactions of the two Magellanic Clouds, occurred about 200-300 Myr ago which are believed to be mainly responsible for the formation of the MB.

  Further study of RR Lyrae stars in the MB and outer regions of the LMC and SMC using near-infrared photometry  will provide valuable information to understand the interaction history of the MS.

\section*{Acknowledgements}
Support for this research has been provided by PRIN INAF 2014 (EXCALIBURS, PI G. Clementini).
We thank the Cambridge Astronomy Survey Unit (CASU) and the Wide Field Astronomy
Unit (WFAU) in Edinburgh for providing calibrated data products under the support of
the Science and Technology Facility Council (STFC) in the UK.
S.S acknowledges research funding support from Chinese Postdoctoral Science Foundation (grant number 2016M590013). 
M.-R.C. acknowledges support from STFC (grant number ST/M00108/1) and from the European Research Council (ERC) under the European Union's Horizon 2020 research and innovation programme (grant agreement No 682115).
R. d. G. acknowledges financial support from the National Natural Science Foundation of China through grants 11373010, 11633005 and U1631102.

\begin{landscape}
 \begin{table}
  \caption{Properties of the  2997 RRab stars in the SMC analysed in the present paper: (1) VMC identification; (2) OGLE identification; (3) Number of the VMC tile; (4), (5) $V$ and $I$ magnitudes from the OGLE~IV catalogue; (6) Pulsation period from OGLE~IV; (7) Dereddened  $K_{\rm s}$ magnitude; (8) Amplitude in $K_{\rm s}$; (9) Reddening; (10) Metallicity from \citet{Skowron2016} transformed to the metallicity scale defined by \citet{Grat2004}; (11) Distance modulus.}
  \label{tab:gen}
  \begin{tabular}{ccccccccccc}
    \hline
VMC id &   OGLE id &Tile &  $V$ & $I$ & Period  & $K_{\rm s,0}$ & $Amp(K_{\rm s})$ &  $E(V-I)$ &  ${\rm [Fe/H]}$ & $(m-M)_0$\\
{} & {} & {} & (mag) & (mag) & (days) & (mag) & (mag) & (mag) & (dex) & (mag)  \\ 
\hline
\hline
J010326.45$-$712127.6  &  OGLE$-$SMC$-$RRLYR$-$1768 &  6\_4 & 20.095 &19.671	&  0.3742896	&    $19.091\pm0.105$  &  0.212  &   $0.03\pm  0.05$  &     $-1.790  \pm  0.330 $  & $19.039   \pm    0.196$ \\	
J012415.27$-$721623.2  &  OGLE$-$SMC$-$RRLYR$-$5285 &  5\_5 & 19.082 &18.675	&  0.3939063	&    $18.296\pm0.049$  &  0.230  &   $-0.04\pm 0.04$  &     $-0.557 \pm  0.042$  & $18.268   \pm    0.119$ \\	
J004801.51$-$733021.8  &  OGLE$-$SMC$-$RRLYR$-$0931 &  4\_3 & 19.813 &19.502	&  0.3995731	&    $18.820\pm0.052$  &  0.163  &   $-0.10\pm  0.04$ &     $-1.790  \pm  0.330 $  & $18.846   \pm    0.169$ \\	
J005300.26$-$725137.0  &  OGLE$-$SMC$-$RRLYR$-$1218 &  4\_3 & 19.965 &19.554	&  0.4035819	&    $19.132\pm0.077$  &  0.268  &   $0.01\pm  0.04$  &     $-1.790  \pm  0.330 $  & $19.170    \pm    0.178$ \\	
J003841.18$-$734423.1  &  OGLE$-$SMC$-$RRLYR$-$0492 &  3\_3 & 19.988 &19.573	&  0.4105743	&    $19.083\pm0.081$  &  0.170  &   $-0.01\pm 0.04$  &     $-1.790  \pm  0.330$  & $19.142   \pm    0.178$ \\	
J003727.81$-$731455.3  &  OGLE$-$SMC$-$RRLYR$-$0432 &  4\_3 & 19.949 &19.509	&  0.4126961	&    $19.113\pm0.069$  &  0.405  &   $0.02\pm  0.04$   &     $-1.790  \pm  0.330 $  & $19.178   \pm    0.173$ \\	
J005845.71$-$733438.3  &  OGLE$-$SMC$-$RRLYR$-$1543 &  4\_4 & 20.245 &19.768	&  0.4127270	&    $19.335\pm0.112$  &  0.484  &   $0.06\pm  0.04$   &     $-1.790  \pm  0.330 $  & $19.400     \pm    0.194$ \\	
J005728.85$-$723454.9  &  OGLE$-$SMC$-$RRLYR$-$1487 &  4\_4 & 20.317 &19.794	&  0.4162569	&    $19.328\pm0.103$  &  0.198  &   $0.09\pm  0.04$   &     $-1.790  \pm  0.330 $  & $19.403   \pm    0.188$ \\	
J010500.09$-$710529.1  &  OGLE$-$SMC$-$RRLYR$-$4686 &  6\_4 & 19.848 &19.340	&  0.4165487	&    $18.844\pm0.066$  &  0.114  &   $0.05\pm  0.04$  &     $-1.790 \pm  0.330 $  & $18.920    \pm    0.171$ \\	
J010412.98$-$732541.8  &  OGLE$-$SMC$-$RRLYR$-$1800 &  4\_4 & 20.158 &19.614	&  0.4182084	&    $19.118\pm0.109$  &  0.157  &   $0.13\pm  0.04$  &     $-1.004 \pm  0.032$  & $19.174   \pm    0.161$ \\	
J005922.20$-$714625.5  &  OGLE$-$SMC$-$RRLYR$-$1581 &  5\_4 & 19.852 &19.418	&  0.4210488	&    $19.006\pm0.047$  &  0.491  &   $0.01\pm  0.04$  &     $-1.118 \pm  0.029$  & $19.074   \pm    0.131$ \\	
J010215.39$-$741252.4  &  OGLE$-$SMC$-$RRLYR$-$1697 &  3\_4 & $-$    &19.580	&  0.423906	&    $18.952\pm0.124$  &  0.136  &   $0.06\pm  0.06$  &     $-1.790  \pm  0.330 $  & $19.048   \pm    0.200  $ \\	
J004639.17$-$731325.2  &  OGLE$-$SMC$-$RRLYR$-$0862 &  4\_3 & 20.067 &19.499	&  0.4243253	&    $18.435\pm0.042$  &  0.122  &   $0.15\pm  0.04$  &     $-1.790  \pm  0.330 $  & $18.532   \pm    0.162$ \\	
J013735.79$-$744047.1  &  OGLE$-$SMC$-$RRLYR$-$5749 &  2\_5 & 19.506 &19.128	&  0.4264885	&    $18.545\pm0.084$  &  0.321  &   $-0.06\pm 0.04$  &     $-1.790  \pm  0.330 $  & $18.648   \pm    0.177$ \\	
J005110.49$-$730750.4  &  OGLE$-$SMC$-$RRLYR$-$1103 &  4\_3 & 20.036 &19.611	&  0.4317203	&    $19.493\pm0.090$  &  0.465  &   $0.01\pm  0.04$  &     $-1.790  \pm  0.330 $  & $19.611   \pm    0.180 $ \\	
J010134.36$-$725427.7  &  OGLE$-$SMC$-$RRLYR$-$1677 &  4\_4 & 19.892 &19.373	&  0.4335109	&    $18.890\pm0.112$  &  0.200   &   $0.06\pm  0.04$  &     $-1.790  \pm  0.330 $  & $19.013   \pm    0.191$ \\	
J005719.65$-$725541.0  &  OGLE$-$SMC$-$RRLYR$-$1476 &  4\_4 & 20.040 &19.593	&  0.4337982	&    $19.122\pm0.120$  &  0.316  &   $0     \pm0.04$     &     $-1.790  \pm  0.330 $  & $19.246   \pm    0.196$ \\	
J005126.22$-$715328.5  &  OGLE$-$SMC$-$RRLYR$-$1117 &  5\_3 & 19.981 &19.496	&  0.4344378	&    $18.990\pm0.094$  &  0.414  &   $0.06\pm  0.04$  &     $-1.790  \pm  0.330 $  & $19.116   \pm    0.181$ \\	
J003257.55$-$732906.3  &  OGLE$-$SMC$-$RRLYR$-$0276 &  4\_2 & 20.169 &19.688	&  0.4362651	&    $19.100\pm0.126$  &  0.433  &   $0.05\pm  0.04$  &     $-1.790  \pm  0.330 $  & $19.230    \pm    0.199$ \\	
J010050.90$-$742433.7  &  OGLE$-$SMC$-$RRLYR$-$4506 &  3\_4 & 19.962 &19.431	&  0.4370365	&    $18.751\pm0.070$  &  0.241  &   $0.10   \pm0.04$  &     $-1.130  \pm  0.041$  & $18.864   \pm    0.139$ \\	
J003618.99$-$725748.0  &  OGLE$-$SMC$-$RRLYR$-$0383 &  4\_3 & 19.892 &19.444	&  0.4373313	&    $18.942\pm0.102$  &  0.461  &   $0.01\pm  0.04$  &     $-1.790  \pm  0.330 $  & $19.075   \pm    0.185$ \\	
J003229.44$-$731557.6  &  OGLE$-$SMC$-$RRLYR$-$0262 &  4\_2 & 19.604 &19.316	&  0.4379744	&    $19.081\pm0.093$  &  0.361  &   $-0.18\pm 0.04$  &     $-0.992 \pm  0.048$  & $19.192   \pm    0.147$ \\	
J010814.34$-$735314.6  &  OGLE$-$SMC$-$RRLYR$-$1975 &  3\_4 & 19.964 &19.461	&  0.4420332	&    $18.972\pm0.104$  &  0.139  &   $0.07\pm  0.04$  &     $-1.790  \pm  0.330 $  & $19.118   \pm    0.186$ \\	
J003418.32$-$724237.9  &  OGLE$-$SMC$-$RRLYR$-$0321 &  4\_2 & 20.085 &19.588	&  0.4451769	&    $19.064\pm0.106$  &  0.380  &   $0.06\pm  0.04$  &     $-1.790  \pm  0.330 $  & $19.219   \pm    0.187$ \\	
J005646.18$-$723452.4  &  OGLE$-$SMC$-$RRLYR$-$1440 &  5\_4 & 20.031 &19.516	&  0.4459867	&    $19.105\pm0.159$  &  0.343  &   $0.08\pm  0.04$  &     $-1.230  \pm  0.032$  & $19.245   \pm    0.202$ \\	
J010653.14$-$743637.7  &  OGLE$-$SMC$-$RRLYR$-$4745 &  3\_4 & 19.802 &19.348	&  0.4484228	&    $18.661\pm0.076$  &  0.386  &   $0.03\pm  0.04$  &     $-1.199 \pm  0.032$  & $18.806   \pm    0.143$ \\	
J005509.35$-$735505.2  &  OGLE$-$SMC$-$RRLYR$-$1358 &  3\_3 & 19.870 &19.399	&  0.4491077	&    $18.949\pm0.065$  &  0.262  &   $0.02\pm  0.04$  &     $-1.790  \pm  0.330 $  & $19.114   \pm    0.166$ \\	
J005327.84$-$733656.3  &  OGLE$-$SMC$-$RRLYR$-$1247 &  4\_3 & 20.354 &19.818	&  0.4497006	&    $19.073\pm0.100$  &  0.332  &   $0.09\pm  0.04$  &     $-1.790  \pm  0.330 $  & $19.239   \pm    0.183$ \\	
J005629.88$-$710719.3  &  OGLE$-$SMC$-$RRLYR$-$4342 &  6\_4 & 19.879 &19.393	&  0.4497817	&    $18.901\pm0.120$  &  0.588  &   $0.03\pm  0.04$  &     $-1.790  \pm  0.330 $  & $19.068   \pm    0.194$ \\	
J004758.96$-$732241.7  &  OGLE$-$SMC$-$RRLYR$-$0928 &  4\_3 & 20.400 &19.898	&  0.4497919	&    $19.341\pm0.163$  &  0.162  &   $0.07\pm  0.04$  &     $-1.790  \pm  0.330 $  & $19.508   \pm    0.224$ \\	
J010535.93$-$720621.9  &  OGLE$-$SMC$-$RRLYR$-$1867 &  5\_4 & 19.932 &19.503	&  0.4538355	&    $18.873\pm0.087$  &  0.278  &   $-0.02\pm 0.04$  &     $-1.790  \pm  0.330 $  & $19.050    \pm    0.175$ \\	
J003057.78$-$735500.2  &  OGLE$-$SMC$-$RRLYR$-$0216 &  3\_2 & 19.765 &19.379	&  0.4545178	&    $18.910\pm0.059$  &  0.418  &   $-0.07\pm 0.04$  &     $-1.241 \pm  0.041$  & $19.072   \pm    0.136$ \\	
J010516.58$-$722526.8  &  OGLE$-$SMC$-$RRLYR$-$1854 &  5\_4 & 20.087 &19.525	&  0.4559563	&    $19.105\pm0.044$  &  0.276  &   $0.12\pm  0.04$  &     $-1.083 \pm  0.033$  & $19.266   \pm    0.123$ \\	
J003651.40$-$713108.5  &  OGLE$-$SMC$-$RRLYR$-$3606 &  5\_3 & 19.913 &19.455	&  0.4563110	&    $18.859\pm0.138$  &  0.182  &   $0.01\pm  0.04$  &     $-1.790  \pm  0.330 $  & $19.043   \pm    0.205$ \\	
J011942.83$-$742418.6  &  OGLE$-$SMC$-$RRLYR$-$5163 &  3\_5 & 19.771 &19.224	&  0.4576531	&    $18.422\pm0.080$  &  0.206  &   $0.11\pm  0.04$  &     $-1.217 \pm  0.029$  & $18.592   \pm    0.145$ \\	
 
 \hline
  \end{tabular}
 \end{table}
The table is published in its entirety in the electronic version of the paper. A portion is shown here for guidance regarding its format and content.
\end{landscape}










\bsp	
\label{lastpage}
\end{document}